\documentclass[final,review,3p,12pt,round,authoryear,times]{elsarticle}

\usepackage{amssymb}
\usepackage{graphicx}
\usepackage{subfig}
\graphicspath{ {./figs/} }
\usepackage{booktabs}
\usepackage{tabularx}
\usepackage{breqn}
\usepackage{amsmath}
\usepackage{lineno}
\usepackage{enumitem}
\usepackage{float}
\usepackage{hyperref} 
\usepackage{tikz}
\usepackage{xcolor}
\usepackage{colortbl}
\usepackage[figurename=Fig.]{caption}

\usepackage{xpatch}
\usepackage{times}
\usepackage{nomencl}
\makenomenclature

\renewcommand{\nompreamble}{The next list describes several symbols that will be later used within the body of the document}
\usepackage{algorithm,algcompatible}
\algnewcommand\INPUT{\item[\textbf{Input:}]}%
\algnewcommand\OUTPUT{\item[\textbf{Output:}]}%

\begin{document}

\begin{frontmatter}

\title{Computational Seismic Fracture Synthesis of Tidal Barrage using Enhanced Isotropic Plasticity Damage Mechanics and Coupled Lagrangian-Eulerian Multiphase Interaction}
\author[Sayan Chowdhury]{Sayan Chowdhury \corref{cor1}}
\cortext[cor1]{Corresponding author at: Marine Autonomous Vessels Laboratory, Indian Institute of Technology Madras, Chennai, Tamil Nadu - 600036, India}
\address[Sayan Chowdhury]{Department of Ocean Engineering, Indian Institute of Technology Madras, Chennai-600036, India}
\ead{sayan.iitm.research@onmail.com}
\author[satya]{Satya Kiran Raju Alluri}
\address[satya]{Scientist-D, National Centre for Coastal Research, Ministry of  Earth Sciences, Chennai-600100, India}
\author[jaya]{Jayaprakash J}
\address[jaya]{Faculty of Structural \& Geotechnical Engineering, Vellore Institute of Technology, Vellore-632014, India}
\author[teo]{Fang Yenn Teo}
\address[teo]{Faculty of Science \& Engineering, University of Nottingham Malaysia, Semenyih, Selangor, Malaysia}
\author[umashankar]{Umashankar M}
\address[umashankar]{Faculty of Environmental \& Water Resources Engineering, Vellore Institute of Technology, Vellore-632014, India}
\begin{abstract}
Mega-engineered hydraulic structures like dams and barrages are critically sensitive to strong ground motion if constructed within the vicinity of triggered fault lines. Collapse post excessive deformation leads to severe environmental impact. In this study, fracture corresponding to the response of a concrete tidal barrage to strong ground motion is analyzed along with behavioral effects due to reservoir-barrage dynamic interaction. An enhanced version of the plasticity damage mechanical model, which includes effects due to degradation of elastic stiffness of concrete as well as restoration of fracture energy losses is assigned as material behavior. The fluid-structure interaction is solved using an idealized Lagrangian-Eulerian formulation. The proposed improvised numerical formulations are validated against benchmark simulations performed on the Koyna dam situated in Maharashtra, India and the results captured are upto 94\% accurate. Finite element simulation of a tidal barrage is performed using a computationally stable mesh with global grid to length ratio of 4.2. The yield surface captured is elliptical in nature and fracture is observed to be propagating from bottom of gate housing covering upto four nodal integration points. 
\end{abstract}



\end{frontmatter}

\section{The Prelude}
\label{sec1}

Tidal barrages are rare concrete spillway structures located within the structural enclosure of a dam or any hydraulic structure used to control the discharge of excess reservoir water onto the downstream end. Generally, spillways have a reservoir on the upstream and a riverbed on the downstream side whereas tidal barrages are an exception for having tidal variation from the gulf or ocean on the downstream side \citep{Murali2017}. The Gulf of Khambhat Development Project also known as the Kalpasar project is one of the first mega projects in India with an aim to construct the world's largest freshwater reservoir dam within the gulf of Khambhat to address the rising drought and freshwater scarcity scenarios in the coastal vicinity of Gujarat, having a reservoir on the upstream end and tidal variations on the downstream end \citep{SubbaRao2011}. The 60-km long earthen masonry dam connects Aladar on the eastern side and Bhavnagar on the western side of the gulf \citep{Sitharam2017}.
Mega-engineered structures are critically sensitive to strong ground motions and undergo excessive deformation if located within the vicinity of the triggered fault lines. Strong ground motions are the strongest shaking that occurs within an approximate radial epicentral distance of 50 km from the sensitive fault lines \citep{Hanks1981}. The response of a relatively rigid structure to strong ground motion is quite abrupt and more critical than any relatively flexible structure. The concrete tidal barrage enclosed within a long stretch of earthen masonry dam is analogous to a system consisting of rigid links connecting two or more identical flexible elements. Sensitivity to seismicity and resulting behavioral aspects of concrete structures, majorly mega-structures like dams and spillways, have been the subject of extensive research during the last ten to fifteen years concerning the safety of such structures. After the historic 1967 Koynanagar earthquake in India with a magnitude of 6.67 Richter scale creating havoc and misery to the daily livelihood in the vicinity of the dam, engineers and scientists found a reason to address the response of dams to seismic forces. It was then, that people started to rethink the environmental impact it would cause due to the catastrophic failure of such mega structures. Non-linear analysis of the Koyna dam was carried out by \cite{Pal1976} for the very first time. \cite{Chopra1973} studied the crack paths of the Koyna dam using the linear elastic method of analysis. The majority of the concrete tidal barrages or spillways in the operational phase have internal zones of micro-cracks present. These cracked zones are formed due to seismic impact, thermal expansion and contraction, irregular foundation settlements and material properties of the hardened concrete \citep{Bazant1985}.
\begin{figure*}
	\centering
	\includegraphics[scale=0.4]{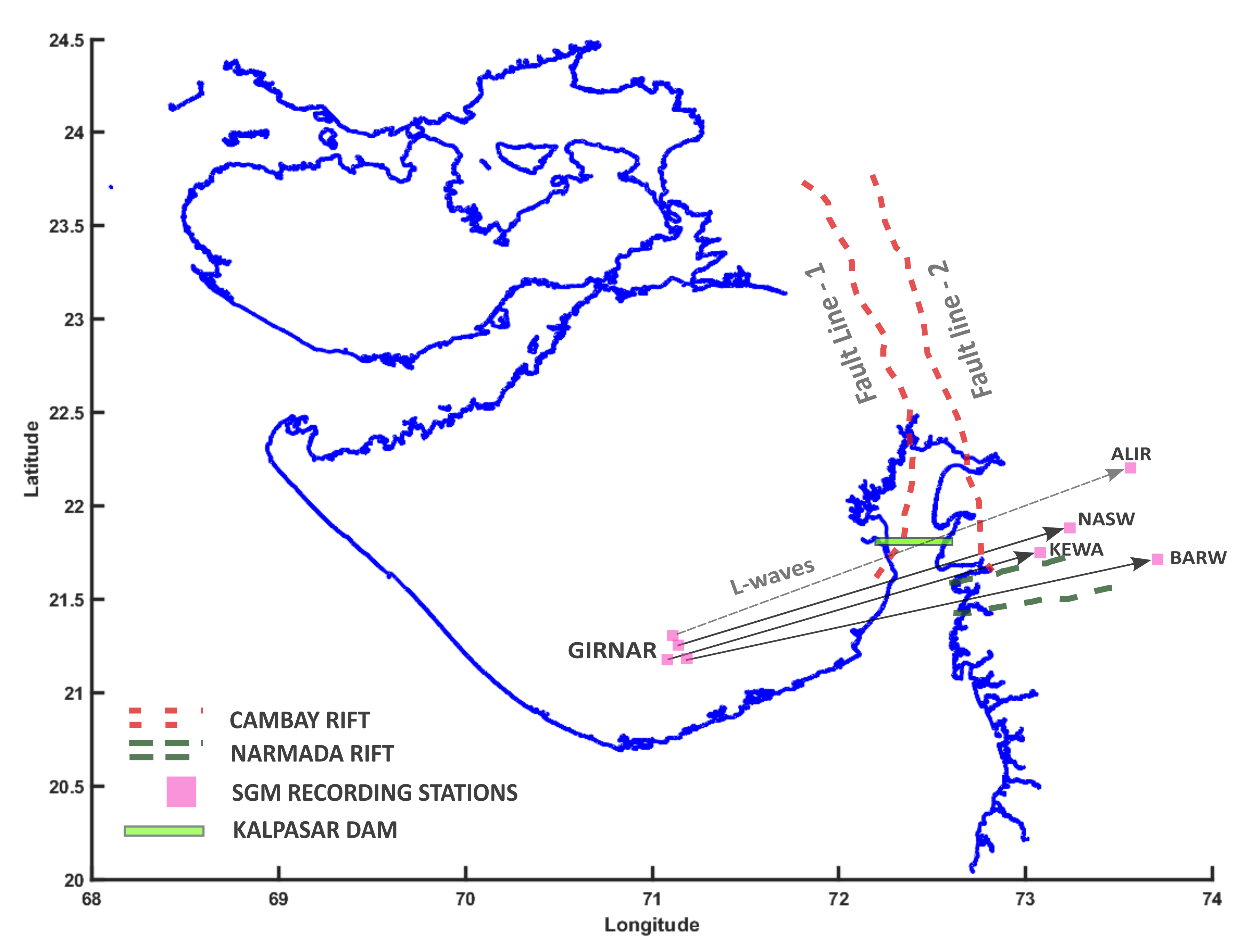}
	\caption{A coastal outline of Gujarat showing traces of historic love waves originating from strong ground motions recorded at stations named GIRNAR, NASW, BARW, ALIR, KEWA with one of he historic path tracing back from station GIRNAR to ALIR passes diagonally through the bottom tectonics of the Kalpasar dam.}
	\label{fig:2}
\end{figure*}
The formation of cracks and micro-cracks due to the structure's response to seismic excitation is catastrophic in nature. Concrete spillways are designed to resist the nominal static earthquake excitation where the material behavior is restricted within the elastic region of the stress-strain curve of the concrete. When the spillway is in the dynamic excitation phase due to maximum credible earthquake forces, the elastic stiffness of the concrete starts degrading slowly and the material enters the strain-softening region of the stress-strain curve. \cite{Hall1988} mentioned in his study that linear analysis of dynamically excited spillway due to strong ground motion can produce critical tensile stresses in the structure exceeding the tensile strength of the structure itself. Quantitative research has been carried out over the last two decades addressing the behavior of fractures and cracks in the concrete due to non-linear responses. The major drawback of the classical theory of linear elastic fracture mechanics is that the non-linear material fracture process, which is significantly large for concrete dams and spillways is neglected. Classical smeared crack analysis from the continuum fracture mechanics theory is less reliable due to the mesh-insensitivity \citep{Cervera2006}. The non-linear material damage configuration of the spillway in the dynamically excited phase can be realistically simulated using computational coupling of damage mechanics and plasticity theory. Damage mechanics is rather an ideology to visualize the material behavior of concrete during the degradation of the elastic stiffness. \cite{Mridha2014} investigated the nonlinear crack propagation in the Koyna dam using the plasticity-damage tension softening model and it was observed that compressive damage was less critical than tensile damage. The state of Gujarat, situated along the western coastal region of the Indian subcontinent is one of the most seismic sensitive intercontinental peninsular regions of the world, susceptible to strong ground motion excitation and critical hazards \citep{Choudhury2018}. Dating back to January 26$^{th}$, 2001, the world witnessed one of the most damaging earthquakes from time immemorial originating at Bhuj, Gulf of Kachchh, India with a magnitude of M$_w$-7.7 on the Richter scale. The majority of the strong ground motions in Gujarat are due to its positional constrain directly on top of the Himalayan collision zone where the Indo-Australian tectonic plate and the Eurasian tectonic plate undergo wedge slipping causing sensitive fault lines at the point of slip \citep{Ni1984}. Subsequent fault lines form a rift in between, two such rifts, the Cambay and the Narmada rift is formed below the gulf of the Khambhat region as shown in \autoref{fig:2}.
The Kalpasar dam as shown in \autoref{fig:2} is critically sensitive owing to its positional constrain. The eastern fault of the Cambay rift as well as the historic love wave path tracing from station GIRNAR to station ALIR falls right underneath the kalpasar dam henceforth increasing its vulnerability to the seismic forces \citep{Chopra2008}.
\section{Enhanced Isotropic plasticity damage mechanical model for material fracture simulation}
The plasticity-damage model is an extensive continuum-based material mechanical process that pragmatically simulates the non-linear behavior of the material post dynamic excitation. The novelty of this theory is that it takes into consideration an associative yield surface having non-associative flow rule, pressure reactivity, flow path reactivity, and is sensitive to both strain hardening and strain softening \citep{Hafezolghorani2018}. The parameter "damage" exclusively defines the non-linear degradation of a material's elastic stiffness after the formation of micro-cracks \citep{Grassl2013}.
\begin{figure*}[!t]
	\centering
	\includegraphics[width=0.9\textwidth, height=6.7cm]{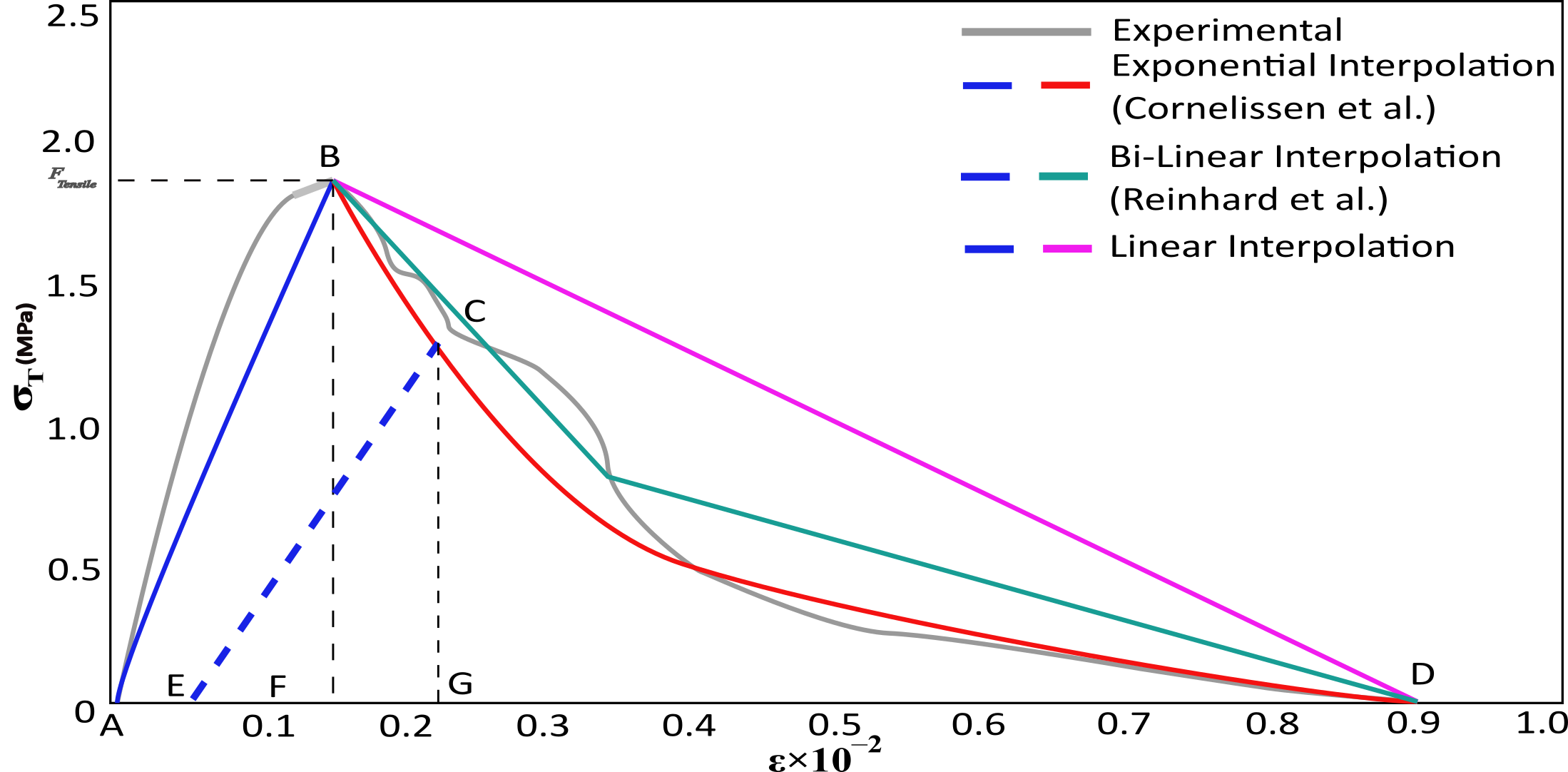}
	\caption{A typical stress-strain curve of a quasi-brittle material under uni-axial tension from split-tensile test comparing \cite{Reinhardt1986a}and \cite{Hillerborg1985} interpolation of the tension softening region.}
	\label{fig:4}
\end{figure*}	
These micro-cracks as already discussed in the prelude are often present as internal deformations within the structural enclosure. Mammoth-scaled Structures like spillways and tidal barrages constructed with quasi-brittle materials are mostly damaged by the formation and propagation of tensile cracks. Split-tensile test using wedge splitting technique was performed by \cite{Bruhwiler1990} on a cylindrical specimens of quasi-brittle material to obtain an experimental stress-strain curve having irregular tension hardening and softening phases as shown in \autoref{fig:4}. The region where the material undergoes strain hardening and tensile stiffness softening, split-tensile test, or rather uni-axial tension test fails to bring out precise results owing to extremely high non-linear deformation rate of the initially present internal micro-crack \citep{kaliharoo}. In \autoref{fig:4}, the curve length from point A to point B is interpolated from erroneous experimental results and represents the elastic region of tensile response. At point B, the material cracks. Post failure the material undergoes strain-softening from point B to C and elastic stiffness degradation from point C to D. At point D, the material is completely fractured. Line EC denotes the unloading-reloading part of the process having slope $\left( {1 - {{\mathord{\buildrel{\lower3pt\hbox{$\scriptscriptstyle\frown$}}\over 
				d} }_{tens}}} \right) \cdot {E_o}$, where ${{{\mathord{\buildrel{\lower3pt\hbox{$\scriptscriptstyle\frown$}}\over 
				d} }_{tens}}}$ is the elastic stiffness degradation of the tension curve and $E_o$ is the elastic modulus. The distance between points A and E is the region of plastic hardening and denoted as $\varepsilon \left( \Im  \right)_{tens}^{^{plastic,h}}$ and the distance between points A and F is the region of plastic cracking denoted by $\varepsilon \left( \Im  \right)_{tens}^{^{cracking,h}}$. The slope of line $\bar E\bar C$ in \autoref{fig:4} is denoted by ${{E_0} \cdot \left( {1 - {{\mathord{\buildrel{\lower3pt\hbox{$\scriptscriptstyle\frown$}}\over 
					d} }_{tens}}} \right)}$ and for line $\bar F\bar C$ it is ${{E_0}}$.			
\subsection{The coupled damage-plasticity formulation}
The directional invariant elasticity and the tensile plasticity are used to formulate the total strain tensor equation \citep{Voyiadjis2008}.
\begin{figure*}
	\centering
	\includegraphics[width=\textwidth]{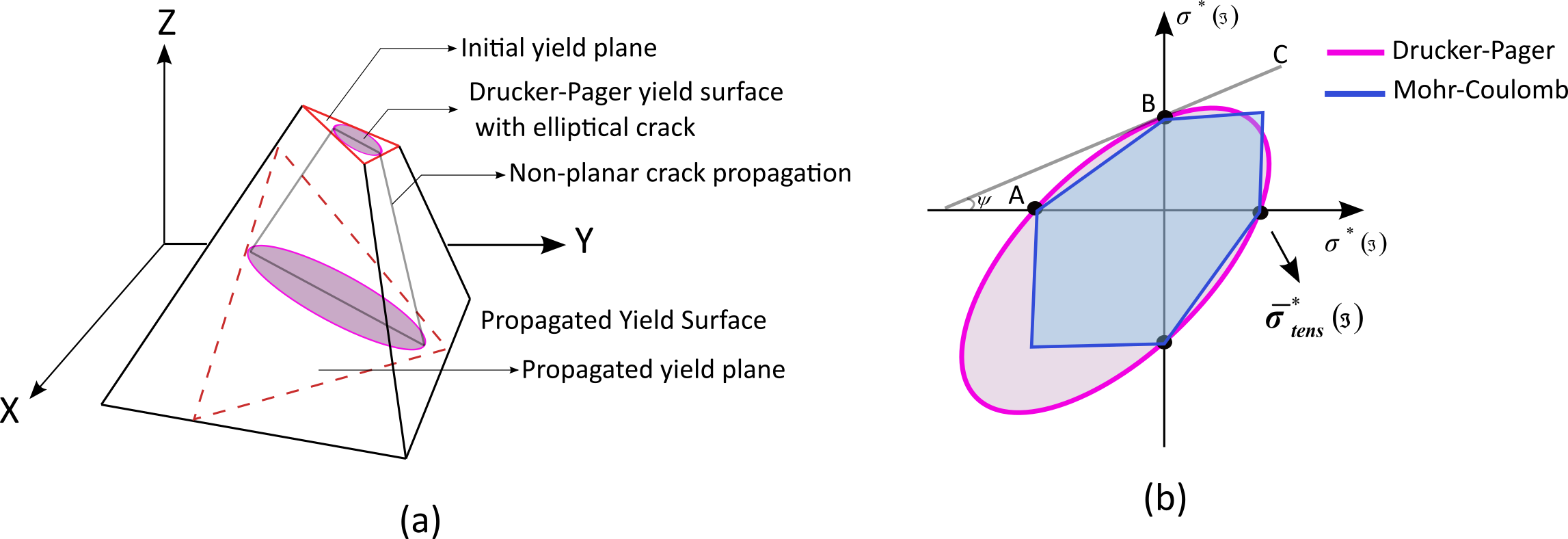}
	\caption{Yield surface and crack insertion zone of a enriched tetrahedral element. (a) shows the propagation of elliptical crack from the initial yield plane to an enriched non-planar yield plane. (b) shows the uniaxial hyperbolic fit of polygonal Mohr-Coulomb and elliptical Drucker-Pager yield profile.}
	\label{fig:5}
\end{figure*}
\begin{linenomath*}
	\begin{equation}
		\sum {\varepsilon \left( \Im  \right)}  = \varepsilon {\left( \Im  \right)^{elastic}} + \varepsilon {\left( \Im  \right)^{plastic}}
	\end{equation}
	\begin{equation}
		\label{eq2}
		{\sigma ^*}\left( \Im  \right) = \mathord{\buildrel{\lower3pt\hbox{$\scriptscriptstyle\frown$}}\over 
			D} {\left( \Im  \right)^{elastic}}:\left( {\varepsilon \left( \Im  \right) - \varepsilon {{\left( \Im  \right)}^{plastic}}} \right)
	\end{equation}
\end{linenomath*}
\begin{linenomath*}
	\begin{equation}
		\label{eq3}
		{{\bar \sigma }^*}\left( \Im  \right) = {{\mathord{\buildrel{\lower3pt\hbox{$\scriptscriptstyle\frown$}}\over 
					D} }_o}{\left( \Im  \right)^{elastic}}:\left( {\varepsilon \left( \Im  \right) - \varepsilon {{\left( \Im  \right)}^{plastic}}} \right)
	\end{equation}
\end{linenomath*}
The nominal stress tensor as given in \autoref{eq2} can be expressed using a scalar damage variable ${\mathord{\buildrel{\lower3pt\hbox{$\scriptscriptstyle\frown$}}\over 
		d} }$ as follows:
\begin{linenomath*}
	\begin{equation}
		\label{eq4}
		{\sigma ^*}\left( \Im  \right) = \left( {1 - \mathord{\buildrel{\lower3pt\hbox{$\scriptscriptstyle\frown$}}\over 
				d} } \right) \cdot {{\mathord{\buildrel{\lower3pt\hbox{$\scriptscriptstyle\frown$}}\over 
					D} }_o}{\left( \Im  \right)^{elastic}}:\left( {\varepsilon \left( \Im  \right) - \varepsilon {{\left( \Im  \right)}^{plastic}}} \right)
	\end{equation}
\end{linenomath*}
\begin{linenomath*}
	\begin{equation}
		\label{eq5}
		{\sigma ^*}\left( \Im  \right) = \left( {1 - {{\mathord{\buildrel{\lower3pt\hbox{$\scriptscriptstyle\frown$}}\over 
						d} }_{tens}}} \right) \cdot {{\bar \sigma }^*}_{tens}\left( \Im  \right) + \left( {1 - {{\mathord{\buildrel{\lower3pt\hbox{$\scriptscriptstyle\frown$}}\over 
						d} }_{comp}}} \right) \cdot {{\bar \sigma }^*}_{comp}\left( \Im  \right)
	\end{equation}
\end{linenomath*}
\begin{linenomath*}
\end{linenomath*}
Since, the effect of compressive damage in mega-structures like spillways, tidal barrages and dams are less critical in nature, the compressive stress tensor becomes negligible and $\left( {{{\bar \sigma }^*}_{comp}\left( \Im  \right) \to 0} \right)$ \citep{Tekie2003}. Therefore rearranging and modifying \autoref{eq5} we get-
\begin{linenomath*}
	\begin{equation}
		\label{eq6}
		{\sigma ^*}{\left( \Im  \right)_{\bmod }} = \left( {1 - {{\mathord{\buildrel{\lower3pt\hbox{$\scriptscriptstyle\frown$}}\over 
						d} }_{tens}}} \right) \cdot {{\bar \sigma }^*}_{tens}\left( \Im  \right)
	\end{equation}
\end{linenomath*}
\begin{linenomath*}
	\begin{equation}
		\left( {1 - {{\mathord{\buildrel{\lower3pt\hbox{$\scriptscriptstyle\frown$}} 
						\over d} }_{tens}}} \right) = \frac{{{\sigma ^*}\left( \Im  \right)}}{{{{\bar \sigma }^*}_{tens}\left( \Im  \right)}}
	\end{equation}
\end{linenomath*}
\begin{linenomath*}
	\begin{equation}
		\label{eq8}
		\therefore {{\mathord{\buildrel{\lower3pt\hbox{$\scriptscriptstyle\frown$}} 
					\over d} }_{tens}} = 1 - \frac{{{\sigma ^*}\left( \Im  \right)}}{{{{\bar \sigma }^*}_{tens}\left( \Im  \right)}}
	\end{equation}
\end{linenomath*}
Where, ${{\mathord{\buildrel{\lower3pt\hbox{$\scriptscriptstyle\frown$}}\over 
			d} }_{tens}}$ is a scalar isotropic tensile damage parameter and ${{\bar \sigma }^*}_{tens}\left( \Im  \right)$ is the uniaxial tensile stress capacity of the quasi-brittle material at failure stage.
Plastic hardening variables with directional invariance were used to control the formation of the yield surface. From the uniaxial tensile test curve as given in \autoref{fig:4} plastic hardening strain was formulated as follows:
\begin{linenomath*}
	\begin{equation}
		\label{eq9}
		\varepsilon {\left( \Im  \right)^{plastic}} = h\left( {\varepsilon {{\left( \Im  \right)}^{plastic,h}},{{\bar \sigma }^*}\left( \Im  \right)} \right) \cdot \sum {\tilde \varepsilon \left( \Im  \right)} 
	\end{equation}
\end{linenomath*}
Neglecting flow in the elastic region of the stress-strain curve given in \autoref{fig:3} and neglecting compressive damage, \autoref{eq9} can be formulated as:
\begin{linenomath*}
	\begin{equation}
		\label{eq10}
		\varepsilon {\left( \Im  \right)^{plastic}} = h\left( {\varepsilon \left( \Im  \right)_{tens}^{^{plastic,h}},{{\bar \sigma }^*}\left( \Im  \right)} \right) \cdot \tilde \varepsilon {\left( \Im  \right)^{plastic}}
	\end{equation}
\end{linenomath*}
$\tilde \varepsilon {\left( \Im  \right)^{plastic}}$ is the governing flow rule in the plastic deformation zone of the yield surface and is given as:
\begin{linenomath*}
	\begin{multline}
		\label{11}
		\tilde \varepsilon {\left( \Im  \right)^{plastic}} = \frac{{\hat \lambda  \cdot \partial }}{{\partial {{\bar \sigma }^*}_{tens}\left( \Im  \right)}}{\left\{ {\hat e \cdot {{\bar \sigma }^*}_{tens}\left( \Im  \right) \cdot {{\tan }^2}\left( \psi  \right)} \right\}^{0.5}}+ \frac{{\hat \lambda  \cdot \partial }}{{\partial {{\bar \sigma }^*}_{tens}\left( \Im  \right)}}{\left\{ {{{\mathord{\buildrel{\lower3pt\hbox{$\scriptscriptstyle\frown$}} 
							\over q} }^2}} \right\}^{0.5}} - \frac{{\hat \lambda  \cdot \partial }}{{\partial {{\bar \sigma }^*}_{tens}\left( \Im  \right)}}\mathord{\buildrel{\lower3pt\hbox{$\scriptscriptstyle\frown$}} 
			\over p}  \cdot \tan \left( \psi  \right)
	\end{multline}
\end{linenomath*}
where, $\mathord{\buildrel{\lower3pt\hbox{$\scriptscriptstyle\frown$}}\over 
	p}  =  - 0.33 \cdot tr\left\{ {{{\bar \sigma }^*}_{tens}\left( \Im  \right)} \right\}$, ${{\mathord{\buildrel{\lower3pt\hbox{$\scriptscriptstyle\frown$}}\over 
			q} }^2} = {\left\{ {{3 \over 2} \cdot {{\bar \sigma }^*}_{deviator}\left( \Im  \right)} \right\}^{0.5}}$, ${\hat e}$ is the flow eccentricity and $\psi $ is the dilation angle. In this study, a fully enriched unstructured tetrahedral element is considered for fracture analysis and a near-elliptical crack is inserted in the yield surface to study the fracture propagation patterns as shown in \autoref{fig:5}. This near-elliptical yield surface is governed by Drucker-Pager's failure theory \citep{Chang1997}. A best fit uniaxial hyperbolic curve from Mohr-Coulomb and Drucker-Pager profiles is formulated as follows:

\begin{linenomath*}
	\begin{multline}
		\label{eqs313}
		{\Upsilon _{hyperbolic}} = \frac{1}{{(1 - {\lambda _1})}} \cdot \mathord{\buildrel{\lower3pt\hbox{$\scriptscriptstyle\frown$}} 
			\over q}  - \frac{1}{{(1 - {\lambda _1})}} \cdot 3\mathord{\buildrel{\lower3pt\hbox{$\scriptscriptstyle\frown$}} 
			\over p} {\lambda _1} + \frac{1}{{(1 - {\lambda _1})}} \cdot {\lambda _2}{{\bar \sigma }^*}{\left( \Im  \right)_{\max }}\varepsilon \left( \Im  \right)_{tens}^{^{plastic,h}} \\- \frac{1}{{(1 - {\lambda _1})}} \cdot {\lambda _3}{{\bar \sigma }^*}{\left( \Im  \right)_{max}}
	\end{multline}
\end{linenomath*}
Where, ${\lambda _1} = {{{\raise0.7ex\hbox{${{{\bar \sigma }^*}_{b0}\left( \Im  \right)}$} \!\mathord{\left/
				{\vphantom {{{{\bar \sigma }^*}_{b0}\left( \Im  \right)} {{{\bar \sigma }^*}_{c0}\left( \Im  \right)}}}\right.\kern-\nulldelimiterspace}
			\!\lower0.7ex\hbox{${{{\bar \sigma }^*}_{c0}\left( \Im  \right)}$}} - 1} \over {{\raise0.7ex\hbox{${2{{\bar \sigma }^*}_{b0}\left( \Im  \right)}$} \!\mathord{\left/
				{\vphantom {{2{{\bar \sigma }^*}_{b0}\left( \Im  \right)} {{{\bar \sigma }^*}_{c0}\left( \Im  \right)}}}\right.\kern-\nulldelimiterspace}
			\!\lower0.7ex\hbox{${{{\bar \sigma }^*}_{c0}\left( \Im  \right)}$}} - 1}}$, ${\lambda _2} = {{{{\bar \sigma }^*}_{comp}\left( \Im  \right)\tilde \varepsilon \left( \Im  \right)_{comp}^{plastic}} \over {{{\bar \sigma }^*}_{tens}\left( \Im  \right)\tilde \varepsilon \left( \Im  \right)_{tens}^{plastic}}} \cdot \left( {1 - {\lambda _1}} \right) - (1 + {\lambda _1})$ and ${\lambda _3} = \frac{{3\left( {1 - {\kappa _{inv}}} \right)}}{{2{\kappa _{inv}}}} - 1$.\\\\
From \autoref{eq8} and \autoref{eq10}, the uniaxial tensile behavior of quasi-brittle material is derived taking into consideration that in tension, failure occurs through crack propagation in the plastic hardening zone.
The equation is formulated as follows:
\begin{linenomath*}
	\begin{equation}
		\label{eq13}
		\varepsilon \left( \Im  \right)_{tens}^{^{plastic,h}} = \varepsilon \left( \Im  \right)_{tens}^{^{cracking,h}} - {{{{\bar \sigma }^*}_{tens}\left( \Im  \right) \cdot {{\mathord{\buildrel{\lower3pt\hbox{$\scriptscriptstyle\frown$}}\over 
							d} }_{tens}}} \over {{E_0} \cdot \left( {1 - {{\mathord{\buildrel{\lower3pt\hbox{$\scriptscriptstyle\frown$}}\over 
								d} }_{tens}}} \right)}}
	\end{equation}
	For quasi-brittle materials like concrete which are devoid of any kind of reinforcements, the response to uniaxial tensile behavior given in \autoref{eq13} is modified by utilizing a stress-displacement curve instead of a stress-strain curve \citep{Yazdani1990}. The equation can be written as follows:
	\begin{linenomath*}
		\begin{equation}
			\label{eq144}
			\delta \left( \Im  \right)_{tens}^{^{plastic,h}} = \delta \left( \Im  \right)_{tens}^{^{cracking,h}} - {{{L_o}{{\bar \sigma }^*}_{tens}\left( \Im  \right) \cdot {{\mathord{\buildrel{\lower3pt\hbox{$\scriptscriptstyle\frown$}}\over 
								d} }_{tens}}} \over {{E_0} \cdot \left( {1 - {{\mathord{\buildrel{\lower3pt\hbox{$\scriptscriptstyle\frown$}}\over 
									d} }_{tens}}} \right)}}
		\end{equation}
	\end{linenomath*}
\end{linenomath*}
\begin{figure*}
	\centering
	\includegraphics[width=0.9\textwidth,height=7cm]{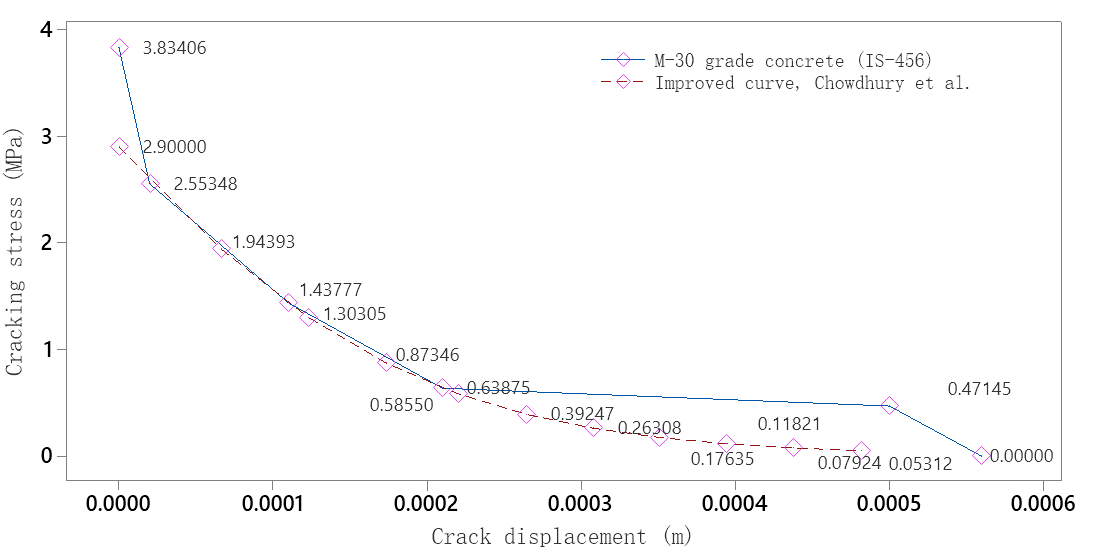}
	\caption{Scattered line plot showing the variation of tensile cracking stress($\delta \left( \Im  \right)_{tens}^{^{cracking,h}}$) with the tensile crack propagation displacement ($\delta \left( \Im  \right)_{tens}^{^{cracking,h}}$), taking into consideration the subsequent tensile strain softening points.}
	\label{fig:6}
\end{figure*}
Where, $\delta \left( \Im  \right)_{tens}^{^{plastic,h}}$ is the material displacement in the plastic flow region under uniaxial split tensile test and ${{L_o}}$ is the standard specimen length. The numerical solution of \autoref{eq144} for M-30 grade Indian concrete mix is presented in form of a graph as shown in \autoref{fig:6}.
\subsection{An enhanced tensile damage-plasticity parameter}
As evident from \autoref{fig:4}, there is a finite curve shift in the tension softening zone of the uniaxial stress-strain tension curve proposed by Reinhard and Cornelissen from the actual experimental curve. This shift is due to losses in fracture energy criteria. Even an infinitesimally small fracture energy loss can result in erroneous values of the elastic stiffness degradation parameter or the tensile damage parameter  ${{{\mathord{\buildrel{\lower3pt\hbox{$\scriptscriptstyle\frown$}} 
				\over d} }_{tens}}}$ leading to formation of coarse fracture paths. This study proposes an enhanced version of the plasticity damage mechanical model by restoring the fracture energy losses back to the computational grid in form of correction factor multiplied with the term ${\sigma ^*}\left( \Im  \right)$ in \autoref{eq8} as follows-			
\begin{linenomath*}
	\begin{equation}
		\label{34}
		{{\mathord{\buildrel{\lower3pt\hbox{$\scriptscriptstyle\frown$}} 
					\over d} }_{tens}} = 1 - \frac{{{\sigma ^*}\left( \Im  \right) \cdot {\gamma_{l}}}}{{{{\bar \sigma }^*}_{tens}\left( \Im  \right)}}
	\end{equation}
\end{linenomath*}
\begin{linenomath*}
	\begin{equation}
		\label{35}
		\therefore {{\mathord{\buildrel{\lower3pt\hbox{$\scriptscriptstyle\frown$}} 
					\over d} }^{imp}}_{tens} = 1 - \frac{{{\sigma ^*}{{\left( \Im  \right)}_{\bmod }}}}{{{{\bar \sigma }^*}_{tens}\left( \Im  \right)}}
	\end{equation}
\end{linenomath*}
\begin{linenomath*}
	\begin{equation}
		\label{35}
		where\;\;\;\;\; {\gamma_{l}} = \oint\limits_{ar} {\mathord{\buildrel{\lower3pt\hbox{$\scriptscriptstyle\frown$}} 
				\over d} _{_{tens}}^{M - 30}}  - \oint\limits_{ar} {\mathord{\buildrel{\lower3pt\hbox{$\scriptscriptstyle\frown$}} 
				\over d} _{_{tens}}^{Interp}}
	\end{equation}
\end{linenomath*}
\begin{linenomath*}
	\begin{multline}
		\label{eq36}
		{\gamma _l} = \oint\limits_{ar} {(8E - 21) \cdot {\sigma ^*}{{\left( \Im  \right)}^3} - (7E - 14) \cdot {\sigma ^*}{{\left( \Im  \right)}^2} \cdot d{\sigma ^*}} - \oint\limits_{ar} { - (3E - 07) \cdot {\sigma ^*}\left( \Im  \right) + 1.00 \cdot d{\sigma ^*}} \\ + \oint\limits_{ar} { - (2E - 07) \cdot {\sigma ^*}\left( \Im  \right) + 1.0059 \cdot d{\sigma ^*}} 
	\end{multline}
\end{linenomath*}
After computation of \autoref{eq36} for ${{\sigma ^*}\left( \Im  \right)}$ ranging from 0.585 MPa to 1.944 MPa, where the curve shift is critical, the fracture energy loss coefficient comes out to be near approximated averaged value of 0.08264.
\section{Computation of coupled multiphase dynamic fluid-structure interaction (CMDFSI)}
A large volumetric mass of freshwater gets confined within the boundary of a dam known as a reservoir and interacts critically with the dam and its components like spillways, and tidal barrages during dynamic excitation due to strong ground motion. The kalpasar dam takes an average inflow of 8000 to 10,000 million cubic meters of water from the major perennial rivers like Sabarmati, Narmada, Mahi, Tapi and Dhadhar \citep{Kolathayar2019}. The inflow of such a huge volumetric quantity results in sloshing of the free water surface. Sloshing can be very well formulated numerically by using the Lagrangian method but due to the combined effects from dynamic excitation of the reservoir bottom as well as the fluid-structure interaction zone, mesh extortions are huge and the Lagrangian method becomes computationally unstable thus an idealized coupled Lagrangian-Eulerian numerical formulation needs to be implemented \citep{Hirt1970}.    
\begin{figure*}
	\centering
	\includegraphics[width=0.9\textwidth]{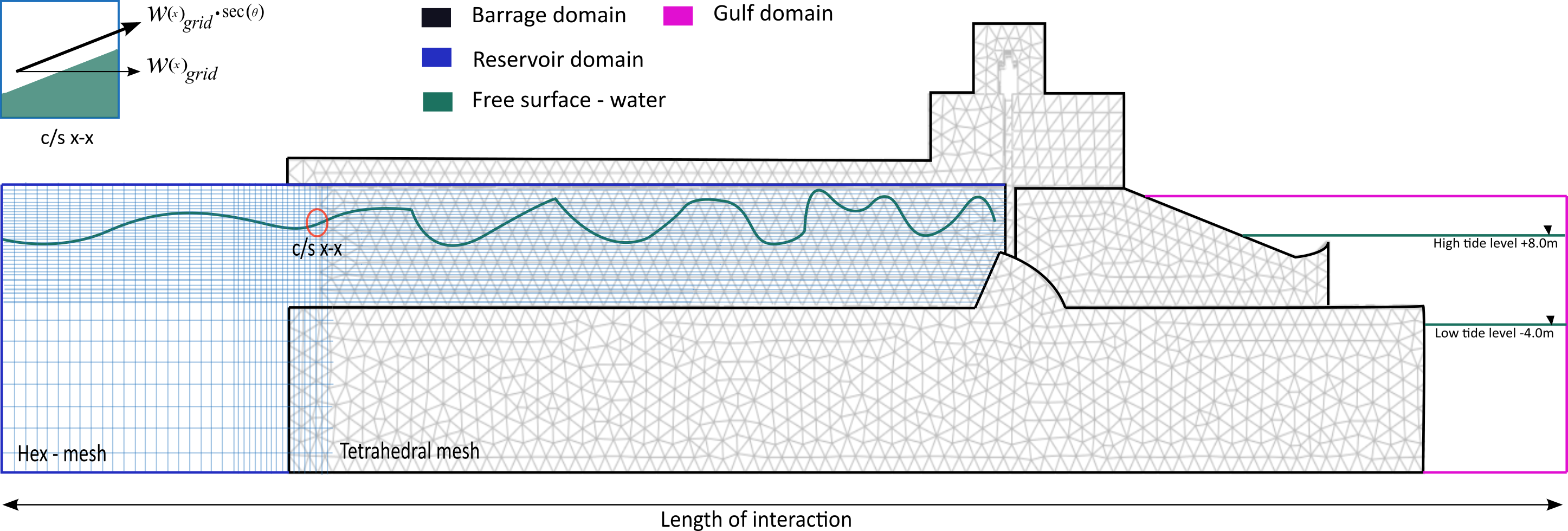}
	\caption{A meshed two dimensional model of barrage highlighting the coupled interaction between the Eulerian-Lagrangian reservoir domain with the structural barrage domain along with velocity induced free water surface sloshing.}
	\label{fig:8}
\end{figure*}
\subsection{Idealized Lagrangian-Eulerian (ILE) formulation}
The generalized form of the famous Navier-Stokes equation with respect to mass, momentum and energy conservation can be written using the divergence theorem as-
\begin{linenomath*}
	\begin{equation}
		\label{eq15}
		\frac{{\partial \rho }}{{\partial t}} + {w_p}(x) \cdot \nabla \rho  + \rho \nabla  \cdot {w_p}(x) = 0{\rm{ }}
	\end{equation}
\end{linenomath*}
\begin{linenomath*}
	\begin{equation}
		\label{eq16}
		\frac{{\partial {w_p}(x)}}{{\partial t}} + {w_p}(x) \cdot \nabla {w_p}(x) + \frac{{\nabla  \cdot P}}{\rho } = 0
	\end{equation}
\end{linenomath*}
\begin{linenomath*}
	\begin{equation}
		\label{eq17}
		\frac{{\partial E(t)}}{{\partial t}} + {w_p}(x) \cdot \nabla E(t) + \frac{P}{\rho }\nabla  \cdot {w_p}(x) = 0{\rm{ }}
	\end{equation}
\end{linenomath*}
Where, $P = \left\{ {(mc\Delta t) - 1} \right\} \cdot \left\{ {\rho E(t) - 0.5\rho {w_p}^2(x)} \right\}$, ${w_p}(x)$ is the velocity of free surface of water in the x-direction of flow, $\rho $ is the mass-density of water at ideal condition and ${E(t)}$ is the energy released or stored.
\autoref{eq15}, \autoref{eq16} and \autoref{eq17} can be written in a generalized form as-
\begin{linenomath*}
	\begin{equation}
		\label{18}
		{\partial  \over {\partial t}}\left( {f_p^*} \right) + {\partial  \over {\partial t}}\Phi \left( {f_r^*} \right) = 0
	\end{equation}
\end{linenomath*}
Where, ${f_p^*}$ is any one fluid property of concern like mass, density, energy etc and $\Phi \left( {f_r^*} \right)$ is the flux of the remaining fluid properties \citep{Chorin1968}.\\
The flux through any flow grid inside the domain of computation can be written as-
\begin{linenomath*}
	\begin{equation}
		\label{eq19} 
		{\Phi _{flux}}\left( {f_p^*} \right) = \int\limits_{\Omega_d} {f_p^* \cdot } {{\vec w}_p}(x) \cdot \hat n \cdot d\Omega_d
	\end{equation}
\end{linenomath*}
Now, the Lagrangian formulation is coupled with \autoref{eq19} to form the so-called Idealized Lagrangian-Eulerian formulation or the ILE-formulation and can be written as-
\begin{linenomath*}
	\begin{equation}
		\label{eq20}
		\Phi _{flux}^{ILE}\left( {f_p^*} \right) = \int\limits_{\Omega _d} {f_p^* \cdot } \left\{ {{{\vec w}_p}(x) - {{\vec w}_{grid}}(x) \cdot \sec (\theta )} \right\} \cdot \hat n \cdot d\Omega_d
	\end{equation}
\end{linenomath*}
Where, ${{{\vec w}_{grid}}(x) \cdot \sec (\theta )}$ is the horizontal component of the grid velocity inside the reservoir which acts as an Eulerian domain as shown in \autoref{fig:8}.
A weighted Galerkin approximation is used to modify \autoref{eq20} by integrating with a weight function ${f_w^{p,2}}$ having polynomial to the order of two with respect to the governing fluid property and another weight function ${{g_w}}$ with respect to the continuity equation \citep{Kuhl2003}. This reduces the computational errors while solving the ILE equation and can be formulated as-
\begin{linenomath*}
	\begin{multline}
		\label{eq21}
		\Phi _{flux}^{ALE,\bmod }\left( {f_p^*} \right) = \int\limits_{\Omega_d} {f_w^{p,2} \cdot f_p^* \cdot } \frac{{\partial {w_p}(x)}}{{\partial t}}\hat n \cdot d\Omega_d + \int\limits_{\Omega_d} {f_w^{p,2} \cdot f_p^*{{\vec w}_p}(x) \cdot \nabla {w_p}(x)\hat n \cdot d\Omega_d} \\ - \int\limits_{\Omega_d} {f_w^{p,2} \cdot f_p^*{{\vec w}_{grid}}(x) \cdot \sec (\theta ) \cdot \nabla {w_p}(x)\hat n \cdot d\Omega_d}  + \int\limits_{\Omega d} {{g_w}} \nabla  \cdot {w_p}(x)\hat n \cdot d\Omega_d
	\end{multline}
\end{linenomath*}
Where, $f_w^{p,2} = \sum\limits_i {{C_1}^i}  \cdot {s_f}^i \cdot \hat n(x) + \sum\limits_i {{C_2}^i}  \cdot {s_f}^i \cdot \hat n(y) + \sum\limits_i {{C_3}^i}  \cdot {s_f}^i \cdot \hat n(z)$, ${g_w} = \sum\limits_i {{C_4}^i}  \cdot {s_f}^i$, $\sum\limits_{n = 1}^4 {C_n^i} $ are arbitrary constants and ${s_f}^i$ is the shape function of the enclosed water volume inside a computational domain.
To compute the solution, we equate $\Phi _{flux}^{ILE,\bmod }\left( {f_p^*} \right) = 0$ as follows-
\begin{linenomath*}
	\begin{multline}
		\label{eq22}
		\int\limits_{\Omega_d} {f_w^{p,2} \cdot } \frac{{\partial {w_p}(x)}}{{\partial t}}\hat n \cdot d\Omega_d + \int\limits_{\Omega_d} {{g_w}} \nabla  \cdot {w_p}(x)\hat n \cdot d\Omega_d + \int\limits_{\Omega_d} {f_w^{p,2} \cdot } {{\vec w}_p}(x)\nabla {w_p}(x)\hat n \cdot d\Omega_d \\- \int\limits_{\Omega_d} {f_w^{p,2} \cdot } {{\vec w}_{grid}}(x) \cdot \sec (\theta )\nabla {w_p}(x)\hat n \cdot d\Omega_d = 0
	\end{multline}
\end{linenomath*}
In coupled Lagrangian-Eulerian technique of solving fluid-structure interaction problems, instead of velocity induced volumetric flow from one computational grid to the other we consider the volumetric water inside a grid to be fixed in temporal space and the grid to be moving from one point to the other within the computational domain \citep{Banks2016}. To address such a technique in the ILE-formulation we need to implement Jacobian interpolation or transformation as shown below-
\begin{linenomath*}
	\begin{equation}
		\label{eq23}
		Jacobian(x,t) = {{dx} \over {d\hat \chi }}\left( {\hat \chi ,t} \right)
	\end{equation}
\end{linenomath*}
Where, $x\left( t \right) \to \hat \chi (t)$ is the computational grid transformation within a spatial-temporal domain. \autoref{eq22} can be re-written in terms of $J(x,t)$ as-
\begin{multline}
	\label{eq23}
	\int\limits_0^{{L_{{\mathop{\rm int}} }}} {\frac{\partial }{{\partial t}}\left\{ {J \cdot {w_p}(x)} \right\} \cdot {f_w}\left( {\hat \chi } \right) \cdot d\hat \chi } + \int\limits_0^{{L_{{\mathop{\rm int}} }}} {J \cdot \frac{\partial }{{\partial x}}{{\vec w}^2}_p(x) \cdot {f_w}\left( {\hat \chi } \right) \cdot d\hat \chi } \\ - \int\limits_0^{{L_{{\mathop{\rm int}} }}} {J \cdot \frac{\partial }{{\partial x}}{{\vec w}_p}(x) \cdot {{\vec w}_{grid}}(x) \cdot \sec (\theta ) \cdot {f_w}\left( {\hat \chi } \right) \cdot d\hat \chi } + \int\limits_0^{{L_{{\mathop{\rm int}} }}} {\frac{\partial }{{\partial x}}\left\{ {{w_p}(x)} \right\} \cdot {g_w}\left( {\hat \chi } \right) \cdot d\hat \chi }  = 0
\end{multline}
\subsection{Lax-Wendroff temporal marching algorithm for ILE computation}
Spatial marching is used in this computational scheme to address the effect of velocity-induced grid transformation over a fixed volume of water within the temporal Eulerian domain. A discretization based on the finite element method is implemented to compute the spatial marching using the Lax-Wendroff algorithm as shown below-
\begin{linenomath*}
	\begin{multline}
		\label{eq25}
		\int\limits_0^{{L_{{\mathop{\rm int}} }}} {{\partial  \over {\partial t}}\left\{ {J \cdot {w_p}^j(x)} \right\} \cdot {f_w}\left( {\hat \chi } \right) \cdot d\hat \chi }  + \int\limits_0^{{L_{{\mathop{\rm int}} }}} {J \cdot {\partial  \over {\partial x}}{{\vec w}^2}{{_p}^j}(x) \cdot {f_w}\left( {\hat \chi } \right) \cdot d\hat \chi } \\ - \int\limits_0^{{L_{{\mathop{\rm int}} }}} {J \cdot {\partial  \over {\partial x}}{w_p}^j(x) \cdot {{\vec w}_{grid}}^j(x) \cdot \sec \left( \theta  \right)}  \cdot {f_w}\left( {\hat \chi } \right) \cdot d\hat \chi  = 0
	\end{multline}
\end{linenomath*}
The continuity equation is neglected since continuity tends to zero towards the boundary of the Lagrangian-Eulerian interaction domain. Using integration by parts for the time-varying element of the transformed grid and then reversing back to the original grid we get-
\begin{linenomath*}
	\begin{multline}
		\label{eq26}
		\int\limits_0^{{L_{{\mathop{\rm int}} }}} {{\partial  \over {\partial t}}\left\{ {J \cdot {w_p}^j(x)} \right\} \cdot {f_w}\left( {\hat \chi } \right) \cdot d\hat \chi }  - \int\limits_0^{{L_{{\mathop{\rm int}} }}} {{\partial  \over {\partial x}}{{\vec w}_p}^j(x) \cdot {f_w}\left( {\hat \chi } \right) \cdot d\hat \chi }  - \int\limits_0^{{L_{{\mathop{\rm int}} }}} {{\partial  \over {\partial x}}{w_p}^j(x) \cdot {{\vec w}_{grid}}^j(x) \cdot \sec \left( \theta  \right)} \\ + \left| {{f_w}\left( {\hat \chi } \right) \cdot {w_p}^j(x)\left\{ {1 - {{\vec w}_{grid}}^j(x) \cdot \sec \left( \theta  \right)} \right\}d\hat \chi } \right|_0^{{L_{{\mathop{\rm int}} }}} = 0 
	\end{multline}
\end{linenomath*}
Performing time integration from ${t^n} \to {t^{n + 1}}$ considering the maximum value of ${\sec \left( \theta  \right)}$ as 1.0 and re-writing the term ${{w_p}^j(x)\left\{ {1 - {{\vec w}_{grid}}^j(x) \cdot \sec \left( \theta  \right)} \right\}}$ as ${{\tilde w}_{ILE}}^j(x)$ we get-
\begin{figure*}[!t]
	\centering
	\includegraphics[width=0.9\textwidth]{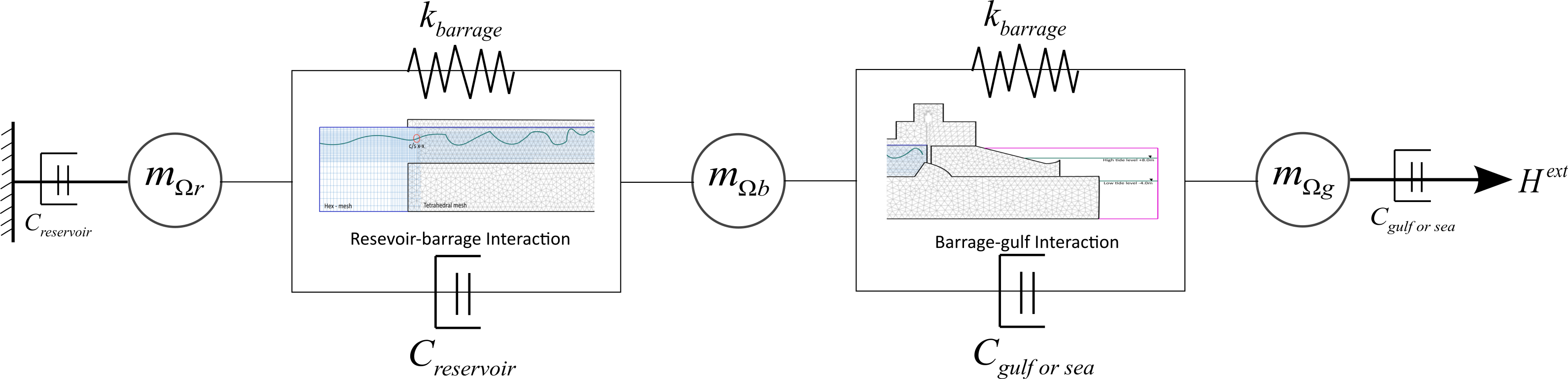}\caption{A modified dash-port model showing the coupling of barrage stiffness with an idealized Lagrangian-Eulerian stiffness proportional damping. The model also depicts the idealization or an assumption of pseudo volumetric water mass, ${{m_{\Omega rd}}}$ inside the reservoir domain for simplified computational convergence.}
	\label{fig9}
\end{figure*}
\begin{linenomath*}
	\begin{multline}
		\label{eq27}
		\int\limits_0^{{L_{{\mathop{\rm int}} }}} {J \cdot {w_p}^j(x){}^{n + 1} \cdot {f_w}\left( {\hat \chi } \right) \cdot d\hat \chi }  - \int\limits_0^{{L_{{\mathop{\rm int}} }}} {J \cdot {w_p}^j(x){}^n \cdot d\hat \chi } - \Delta t\int\limits_0^{{L_{{\mathop{\rm int}} }}} {{\partial  \over {\partial \hat \chi }} \cdot {f_w}\left( {\hat \chi } \right) \cdot {{\tilde w}_{ILE}}^j{{(x)}^{{{n + 1} \over 2}}}}  \cdot d\hat \chi \\ + \Delta t\left| {{f_w}\left( {\hat \chi } \right) \cdot {{\tilde w}_{ILE}}^j{{(x)}^{{{n + 1} \over 2}}}} \right|_0^{{L_{{\mathop{\rm int}} }}} = 0  
	\end{multline}
\end{linenomath*}
Using, spatial finite-element mesh discretization the assembled equation obtained is as follows:
\begin{linenomath*}
	\begin{multline}
		\label{eq28}
		{\left[ {\begin{array}{*{20}{c}}
					{{m_{\Omega b}}}&0&0\\
					0&{{m_{\Omega rd}}}&0\\
					0&0&{{m_{\Omega gd}}}
			\end{array}} \right]^{n + 1}}{\left\{ {{{\dot w}_p}^j(x)} \right\}^{n + 1}} - n \cdot {\left[ {\begin{array}{*{20}{c}}
					{{m_{\Omega b}}}&0&0\\
					0&{{m_{\Omega rd}}}&0\\
					0&0&{{m_{\Omega gd}}}
			\end{array}} \right]^n}{{\dot w}_p}^j(x) - \Delta {\left\{ {{G^j}_{res}} \right\}^{\frac{{n + 1}}{2}}} = 0
	\end{multline}
\end{linenomath*}
Where, ${{m_{\Omega b}}}$ is the mass within the barrage domain, ${{m_{\Omega rd}}}$ is the mass within the ILE domain, ${{m_{\Omega gd}}}$ is the finite mass within the sea-water or gulf domain and ${{G^j}_{res}}$ is the residual matrix within the temporal space calculated at half-step time. Applying the explicit Lax-Wendroff computation scheme, the numerical approximation of \autoref{eq28} can be solved using the following relations-
\begin{linenomath*}
	\begin{multline}
		\label{eq29}
		{{\tilde F}_{LWS}}^{\frac{{j + 1}}{2}}(x) = \frac{1}{2}\left\{ {{w_p}^{j + 1,n}(x) + {w_p}^{j,n}(x)} \right\} + \frac{{\Delta ts}}{{2\Delta x}}\left\{ {{{\tilde F}_{LWS}}({{\dot w}_p}^{j,n}(x))} \right\} - \frac{{\Delta ts}}{{2\Delta x}}\left\{ {{{\tilde F}_{LWS}}({{\dot w}_p}^{j + 1,n}(x))} \right\}
	\end{multline}
\end{linenomath*}
Where,${{\tilde F}_{LWS}}(x)$ is the Lax-Wendroff flux function and ${\Delta ts}$ is the computational stability of the temporal domain which can be formulated in terms of the classical Courant-Friedrichs-Lewy parameter (${P_{CFL}}$) as follows-
\begin{linenomath*}
	\begin{equation}
		\label{eq30}
		\Delta ts = {P_{CFL}} \cdot \min \left\{ {{{{L_{{\mathop{\rm int}} }}} \over {{C_o} + {\mu _d}}}} \right\}
	\end{equation}
\end{linenomath*}
where, ${{C_o}}$ is the local speed of sound in ideal condition and ${{\mu _d}}$ is the dynamic viscosity of the fluid. The value of ${{C_o}}$ is taken as 1500 and the value of ${{\mu _d}}$ is considered as 0.001 \citep{Zeng2012}.
\subsection{Critical damping constant computation using Westergaard's added mass scheme}
During dynamic excitation from strong ground motion, the barrage-reservoir acts as a critical coupled system. The volumetric mass of the water acts as an added mass damper, damping out the overall motion of the system. The critical damping is computed using Westergaard's added mass formulation wherein the Rayleigh's stiffness proportional damping constant is considered. A modified dash-pot system as shown in \autoref{fig9} is sketched to visualize and compute the required critical damping. The critical damping constant required to compute the Idealized Lagrangian-Eulerian coupling based on Westergaard theory can be formulated from Rayleigh's damping ratio formula which is valid within the cut-off frequency range or can be computed with respect to the fundamental frequency of vibration as given below-
\begin{linenomath*}
	\begin{equation}
		\label{31}
		\mathop {\min }\limits_{{\alpha _r}{\beta _r}} \sum\limits_{j = 1}^n {\left\{ {{{{\alpha _r}} \over {2{\varpi _j}}} + {{{\beta _r}{\varpi _j}} \over 2} - {\zeta _r}} \right\}}  = D_{rayleigh}^{critical}
	\end{equation}
\end{linenomath*}
At, \textit{j}=1, where the structure undergoes fundamental mode of vibration the minimum value of $D_{rayleigh}^{critical}$ tends to zero as given \citep{Bea2009}-
\begin{linenomath*}
	\begin{equation}
		\label{eq32}
		\forall j = 1 \;\;\; \mathop {\min }\limits_{{\alpha _r}{\beta _r}} \sum\limits_{j = 1}^n {\left\{ {{{{\alpha _r}} \over {2{\varpi _j}}} + {{{\beta _r}{\varpi _j}} \over 2} - {\zeta _r}} \right\}}  \to 0
	\end{equation}
\end{linenomath*}
Rearranging the above equation with ${\alpha _r} = 0$, ${\varpi _1} = 6.283 * {f_{nat}}$ and ${\zeta _r} = 0.03$ we get-
\begin{linenomath*}
	\begin{equation}
		{\beta _r} = {{2{\zeta _r}} \mathord{\left/
				{\vphantom {{2{\zeta _r}} {{\varpi _j}}}} \right.
				\kern-\nulldelimiterspace} {{\varpi _j}}} = 0.003407
	\end{equation}
\end{linenomath*}
Where, ${\beta _r}$ is Rayleigh's stiffness proportional damping constant, ${\zeta _r}$ is the critical damping ratio and ${\varpi _1}$ is the fundamental frequency of vibration in rads/sec. The idealized pseudo water mass added on the reservoir-barrage interaction side and barrage-gulf interaction side can be computed using-
\begin{linenomath*}
	\begin{equation}
		\label{34}
		{m_{\Omega rd}} = 0.875{\rho _{res}}\left\{ {{h^2}_{{\mathop{\rm int}} ,\Omega rb} - {h_{{\mathop{\rm int}} ,\Omega rb}} \cdot y_{res}^{\max }} \right\}
	\end{equation}
\end{linenomath*}
\begin{linenomath*}
	\begin{equation}
		\label{35}
		{m_{\Omega rg}} = 0.875{\rho _{sea}}\left\{ {{h^2}_{{\mathop{\rm int}} ,\Omega bg} - {h_{{\mathop{\rm int}} ,\Omega bg}} \cdot \left| {{y_{sea}}} \right|_{Low\;tide}^{High\;tide}} \right\}
	\end{equation}
\end{linenomath*}
Where ${\rho _{res}} = 1000\;{\rm{kg/}}{{\rm{m}}^3}$, ${\rho _{sea}} = 1028\;{\rm{kg/}}{{\rm{m}}^3}$, ${{h_{{\mathop{\rm int}} ,\Omega rb}}}$ is the height of interaction between reservoir and barrage, ${{h_{{\mathop{\rm int}} ,\Omega bg}}}$ is the height of interaction between the barrage and gulf, ${y_{res}^{\max }}$ is the maximum reservoir water level and ${\left| {{y_{sea}}} \right|_{Low{\rm{ }}tide}^{High{\rm{ }}tide}}$ is the water level in the gulf corresponding to low and high tide scenarios.
\section{Computation of non-linear dynamic multiphase equation using numerical methods }
An improved version of the time-discrete dynamic equilibrium equation for the barrage-reservoir as well as the barrage-gulf interaction can be formulated as follows-
\begin{linenomath*}
	\begin{multline}
		\left[ M \right]{\left\{ {{a_p}^j(x)} \right\}^{n + 1}} - n \cdot \left[ M \right]{a_p}^j(x) + \left( {1 + {\lambda _1}} \right)\left[ C \right]{\left\{ {{w_p}^j(x)} \right\}^{n + 1}} - {\lambda _1}\left[ C \right]{\left\{ {{w_p}^j(x)} \right\}^n} \\+ \left( {1 + {\lambda _1}} \right)\left[ {{K_b}} \right]{\left\{ {{d_p}^j(x)} \right\}^{n + 1}} - {\lambda _1}\left[ {{K_b}} \right]{\left\{ {{d_p}^j(x)} \right\}^n}= \left( {1 + {\lambda _1}} \right)F_{dyn}^{^{n + 1}} - {\lambda _1}F_{dyn}^{^n} + \Delta {\left\{ {{G^j}_{res}} \right\}^{\frac{{n + 1}}{2}}}
		\label{eq400}
	\end{multline}
\end{linenomath*}
\autoref{eq400} can be re-written using matrix expansion as follows-
\begin{linenomath*}
	\begin{multline}
		\left[ {\begin{array}{*{20}{c}}
				{{m_{\Omega b}}}&0&0\\
				0&{{m_{\Omega rd}}}&0\\
				0&0&{{m_{\Omega gd}}}
		\end{array}} \right]{\left\{ {{{\dot w}_p}^j(x)} \right\}^{n + 1}} - n \cdot \left[ {\begin{array}{*{20}{c}}
				{{m_{\Omega b}}}&0&0\\
				0&{{m_{\Omega rd}}}&0\\
				0&0&{{m_{\Omega gd}}}
		\end{array}} \right]{{\dot w}_p}^j(x) + \left( {1 + {\lambda _1}} \right)\left[ {\begin{array}{*{20}{c}}
				0\\
				{{C_r}}\\
				{{C_g}}
		\end{array}} \right]{\left\{ {{w_p}^j(x)} \right\}^{n + 1}} \\- {\lambda _1}\left[ {\begin{array}{*{20}{c}}
				0\\
				{{C_r}}\\
				{{C_g}}
		\end{array}} \right]{\left\{ {{w_p}^j(x)} \right\}^n} + \left( {1 + {\lambda _1}} \right)\left[ {{K_b}} \right]{\left\{ {\int {{w_p}^j(x) \cdot dt} } \right\}^{n + 1}} - {\lambda _1}\left[ {{K_b}} \right]{\left\{ {\int {{w_p}^j(x) \cdot dt} } \right\}^n}\\ = \left( {1 + {\lambda _1}} \right)F_{dyn}^{^{n + 1}} - {\lambda _1}F_{dyn}^{^n} + \Delta {\left\{ {{G^j}_{res}} \right\}^{\frac{{n + 1}}{2}}}
		\label{eq401}
	\end{multline}
\end{linenomath*}
Where, ${{a_p}^j(x)}$, ${{w_p}^j(x)}$ and ${{d_p}^j(x)}$ are the near approximated acceleration, velocity and displacement respectively at the ${j^{th}}$ node at any time step of ${t^{n + 1}}$. $\left[ M \right]$, $\left[ C \right]$, $\left[ K \right]$ are the mass, damping and stiffness matrices of the coupled fluid-structure ILE system respectively.
\subsection{Corrected implicit-$\lambda$ integration scheme}
Using the Newmark difference scheme ${{w_p}^j(x)}$ and ${{d_p}^j(x)}$ can be expanded as following-
\begin{linenomath*}
	\begin{equation}
		{w_p}^j{(x)^{n + 1}} = {w_p}^j{(x)^n} + \Delta t(1 - {\lambda _3}) \cdot {a_p}^j{(x)^n} + \Delta t{\lambda _3}{a_p}^j{(x)^{n + 1}}
		\label{eq402}
	\end{equation}
\end{linenomath*}
\begin{linenomath*}
	\begin{equation}
		{d_p}^j{(x)^{n + 1}} = {d_p}^j{(x)^n} + \Delta t \cdot {w_p}^j{(x)^n} + 0.5\Delta {t^2}\left\{ {\left( {1 - 2{\lambda _2}} \right){a_p}^j{{(x)}^n} + 2{\lambda _2}{a_p}^j{{(x)}^{n + 1}}} \right\}
		\label{eq403}
	\end{equation}
\end{linenomath*}
\begin{algorithm*}[!t]
	\caption{Predictor-corrector time marching algorithm}
	\begin{algorithmic}[1]
		\STATE \textbf{Predict and update dynamic response vectors}\newline ${{\hat d}_p}^{(1)}{(x)^{n + 1}} = {d_p}^{(1)}{(x)^n} + \Delta t \cdot {w_p}^{(1)}{(x)^n} + 0.5\Delta {t^2}\left( {1 - 2{\lambda _2}} \right){a_p}^{(1)}{(x)^n}$ \newline ${{\hat w}_p}^{(1)}{(x)^{n + 1}} = {w_p}^{(1)}{(x)^n} + \Delta t(1 - {\lambda _3}) \cdot {a_p}^{(1)}{(x)^n}$ \newline ${a_p}^{(1)}{(x)^{n + 1}} = 0$
		\STATE \textbf{Correct predicted vectors for given \boldmath ${a_p}^j{(x)^{n + 1}}$}\newline ${d_p}^{(1)}{(x)^{n + 1}} = {{\hat d}_p}^{(1)}{(x)^{n + 1}} + \Delta {t^2} \cdot {\lambda _2}{a_p}^j{(x)^{n + 1}}$ \Comment{displacement corrector phase for given seismic acceleration}\newline ${w_p}^{(1)}{(x)^{n + 1}} = {{\hat w}_p}^{(1)}{(x)^{n + 1}} + \Delta t{\lambda _3} \cdot {a_p}^j{(x)^{n + 1}}$ \Comment{velocity corrector phase for given seismic acceleration}\newline ${a_p}^{(1)}{(x)^{n + 1}} \ne 0$ 
		\STATE \textbf{Compute the dynamic response component}\newline $\left[ M \right]{\left\{ {{a_p}^j(x)} \right\}^{n + 1}} - n \cdot \left[ M \right]{a_p}^j(x) + \left( {1 + {\lambda _1}} \right)\left[ C \right]{\left\{ {{w_p}^j(x)} \right\}^{n + 1}} - {\lambda _1}\left[ C \right]{\left\{ {{w_p}^j(x)} \right\}^n} + \left( {1 + {\lambda _1}} \right)\left[ {{K_b}} \right]{\left\{ {{d_p}^j(x)} \right\}^{n + 1}} - {\lambda _1}\left[ {{K_b}} \right]{\left\{ {{d_p}^j(x)} \right\}^n}$ \newline set iterative counter flags for \textit{j = 1,2,3....,m} to run multiple corrector loop.
		\STATE \textbf{Compute the residual component}\newline
		${R^{j,n + 1}} = \left( {1 + {\lambda _1}} \right)F_{dyn}^{^{n + 1}} - {\lambda _1}F_{dyn}^{^n} + \Delta {\left\{ {{G^j}_{res}} \right\}^{\frac{{n + 1}}{2}}}$
		\STATE \textbf{Check the convergence criterion \boldmath $\;\; \forall j > 1$}\newline \textbf{If} $\left| {{R^{j,n + 1}}} \right| \le {\partial _{tolerance}}\sum\limits_{j = 1}^n {\left| {{R^{j,n + 1}}} \right|}$ \Comment{convergence achieved} \newline      \textbf{then} update ${d_p}^{j + 1}{(x)^n} \to {d_p}^{j + 1}{(x)^{n + 1}}$ and ${w_p}^{j + 1}{(x)^n} \to {w_p}^{j + 1}{(x)^{n + 1}}$ \newline \textbf{else} return to step-(3) and set counter as \boldmath $j = j + 1$ and repeat
		\label{algo1}
	\end{algorithmic}
\end{algorithm*}
where, $\lambda_1$, $\lambda_2$, $\lambda_3$ are the stability and precision control parameters. To attain unconditional stability and precision up to second order accuracy the value of $\lambda_1$ is taken as -0.166, $\lambda_2$ as 0.339 and $\lambda_3$ as 0.666 using the mathematical relations as follows \citep{Gladwell1980}-
\begin{linenomath*}
	\begin{equation}
		{\lambda _2} = 0.25{\left( {1 - {\lambda _1}} \right)^2}{\rm{\;and\;}}{\lambda _3} = 0.5 - {\lambda _1}
		\label{eq404}
	\end{equation}
\end{linenomath*}
Rearranging \autoref{eq402} and \autoref{eq403} the predictor and corrector phases of the non-linear dynamic equation is written as follows \citep{Diethelm2002}-
\begin{linenomath*}
	\begin{equation}
		{{\hat w}_p}^j{(x)^{n + 1}} = {w_p}^j{(x)^n} + \Delta t(1 - {\lambda _3}) \cdot {a_p}^j{(x)^n}
		\label{eq405}
	\end{equation}
\end{linenomath*}
\begin{linenomath*}
	\begin{equation}
		{w_p}^j{(x)^{n + 1}} = {{\hat w}_p}^j{(x)^{n + 1}} + \Delta t{\lambda _3} \cdot {a_p}^j{(x)^{n + 1}}
		\label{eq407}
	\end{equation}
\end{linenomath*}
\begin{linenomath*}
	\begin{equation}
		{{\hat d}_p}^j{(x)^{n + 1}} = {d_p}^j{(x)^n} + \Delta t \cdot {w_p}^j{(x)^n} + 0.5\Delta {t^2}\left( {1 - 2{\lambda _2}} \right){a_p}^j{(x)^n}
		\label{eq406}
	\end{equation}
\end{linenomath*}
\begin{linenomath*}
	\begin{equation}
		{d_p}^j{(x)^{n + 1}} = {{\hat d}_p}^j{(x)^{n + 1}} + \Delta {t^2} \cdot {\lambda _2}{a_p}^j{(x)^{n + 1}}
		\label{eq407}
	\end{equation}
\end{linenomath*}
Finally, the non linear dynamic equation is computed based on the corrected implicit-$\lambda$ integration scheme using a predictor-corrector time marching algorithm presented above as Algorithm 1.
\section{Computational validation of EIPDM and CMDFSI numerical schemes}
Several significant analyses and research work have been carried out on the Koynanagar dam located in Maharashtra, India. The koynanagar dam is the only largest dam that was exposed to one of the most devastating strong ground shaking dating back to 11\textsuperscript{th} December 1967 with a Richter scale of 6.7-Mw. Extensive cracks were formed on both the downstream and upstream slopes causing critical damage to the structure. A non-linear dynamic fracture simulation was performed on the Koynanagar concrete dam model exposed to the 1967 Koynanagar accelerogram using a coaxial rotational crack model and Lagrangian fluid-structure interaction. The finite element technique was adapted to simulate the two-dimensional model of the Koyna dam. This study proposing an improved version of the damage-plasticity model as the material parameter of concrete and an idealized multiphase Lagrangian-Eulerian fluid-structure interaction model is validated with the benchmark results put forward by \cite{Calayir2005}. The dam foundation interaction is neglected keeping in mind the foundation behaves like a rigid surface with an approximate elastic modulus of 7.22x10\textsuperscript{4} MPa, thus the seismic acceleration has been applied at the interface of the dam and foundation. An Eulerian-Lagrangian single-phase fluid domain is considered with a mass density of 1000 kg/m\textsuperscript{3}, bulk modulus of 2078 MPa, and rotational constrain of 1.6x10\textsuperscript{5} MPa. The speed of sound over the water surface is considered 1570 m/s. The height of interaction between the water and the dam domain is taken as 90 m and the length of interaction for the reservoir domain is considered 210 m. The base of the reservoir is considered as rigid and nodal displacement is taken as zero for the abridged reservoir boundaries and only the free water surface is allowed to undergo sloshing as a part of the idealized single-phase fluid-structure interaction phenomenon. An enhanced isotropic plasticity damage model with tensile damage subroutine is assigned as material behavior of concrete dam.
\begin{figure*}[!t]
	\centering
	\includegraphics[width=0.9\textwidth]{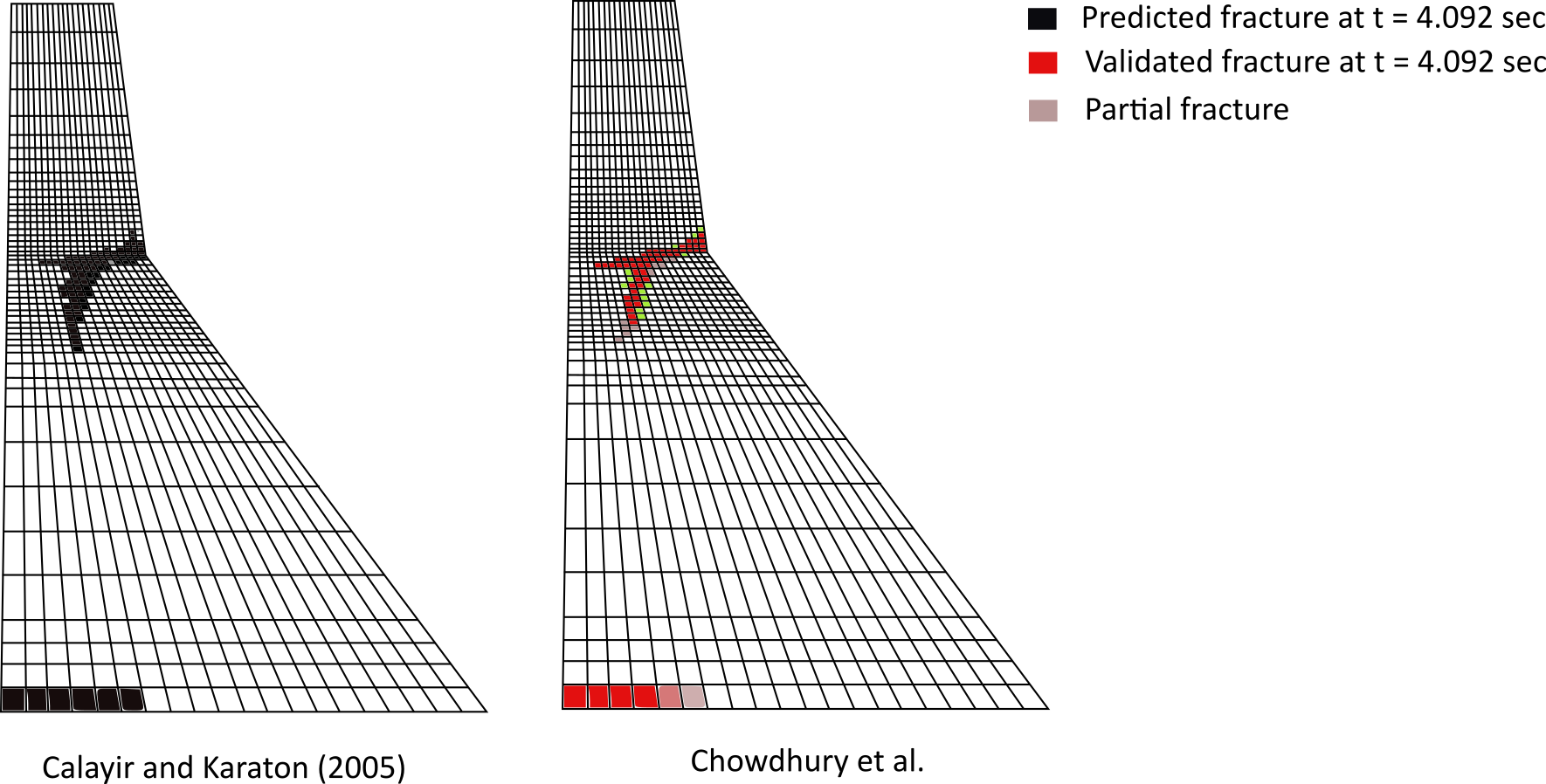}\caption{A two dimensional finite element simulation of the Koyna dam comparing fracture profile obtained using benchmark models and proposed model at a seismic excitation time of 4.092 seconds.}
	\label{fig11}
\end{figure*}
The elastic modulus of the concrete is taken as 27,838 MPa, Poisson ratio as 0.2, mass density as 2670 kg/m\textsuperscript{3} and yield strength in compression as 31 MPa. Dynamic variation of both elastic moduli of the Eulerian domain as well as the structural domain is ignored in the simulation. Rayleigh's stiffness proportional damping is considered with critical damping constant of 0.005 and fundamental frequency of vibration as 3.012 cycles/sec. A structured quad-dominated mesh with a 2.5\% bias ratio is considered for analysis. Results obtained from finite element simulation using Abaqus is validated against benchmark experimental results obtained by Calayir and Karaton (2005). It is observed that the results obtained in this study are well within the validation domain with 94.12\% accuracy. The fracture profile obtained by Calayir et al. spreads over a range of 34 unique nodal points near the upper portion of the downstream monolith whereas the profile obtained by this study spreads over a range of 32 unique nodal points at seismic excitation time of 2.4 seconds. As the simulation proceeds further, fracture starts propagating towards the upstream end, and crack opening in form of integration point deletion takes place. A snapshot of the fracture profile generated on the dam monolith at an excitation time of 4.092 seconds is successfully captured and validated as shown in \autoref{fig11}. 
Results obtained by Calayir et al. captured 80 unique nodal points undergoing damage whereas computations performed in this study captured 74 unique nodal points undergoing damage near the upper portion of the monolith with an accuracy of 92.6\%. The enhanced isotropic plasticity damage mechanical model proposed in this study is able to capture a smooth fracture profile that is computationally stable than those obtained in the benchmark experiment. The Coaxial Rotational Crack model used by Calayir et al. did not consider the losses in fracture energy while interpolating the tension softening behavior of the material and thus could only capture a coarse fracture path. Since fracture energy losses is incorporated with the plasticity damage equation in the present study, even partially fractured element or rather elements that falls under the critical fracture path is smoothly captured and highlighted as shown in \autoref{fig11}. It is observed that material yields at an excitation time, \textit{t=2.2} seconds and fracture propagation initiates. The majority of the critical damage due to fracture is achieved within \textit{t=4.09} seconds of seismic excitation.
\begin{figure*}[!t]
	\centering
	\includegraphics[width=0.9\textwidth, height=7cm]{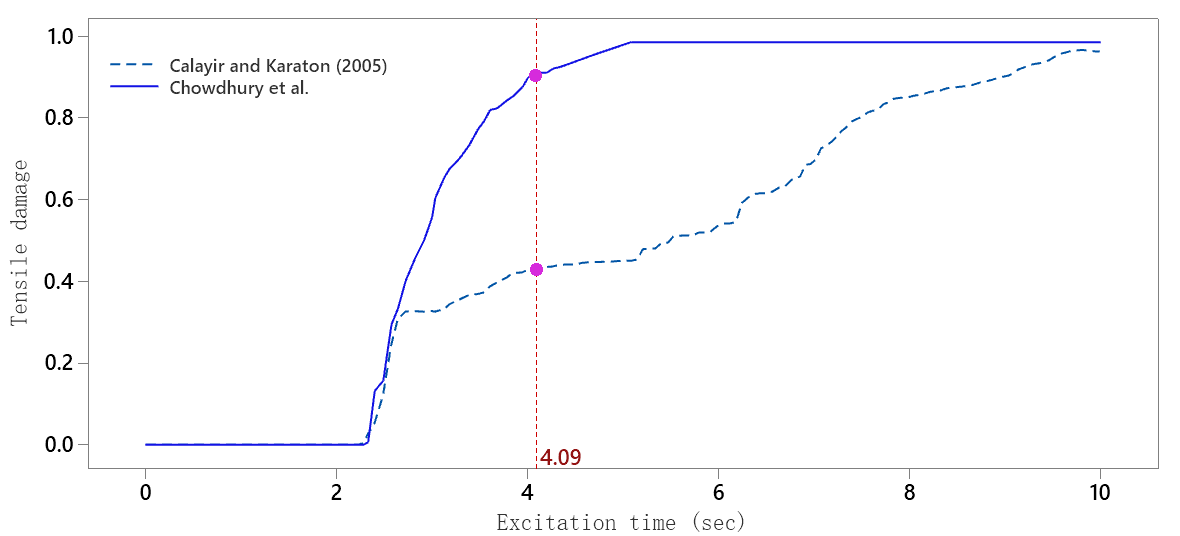}\caption{A tensile damage curve of the Koyna dam plotted against seismic excitation time comparing stable and asymptotic profile obtained using proposed models with benchmark results. }
	\label{fig12}
\end{figure*}
 The tensile damage curve obtained from simulations performed by Calayir et al. showed that fracture propagation was quite unstable both before and after critical damage was achieved whereas simulations performed in the present study obtained a much more realistic and stable damage curve which also has a horizontal asymptotic part post time \textit{t=5.1} seconds as shown in \autoref{fig12}. The damage parameter in any material should not exceed the ideal value of 1.0 which suggests that once the fracture is achieved and it propagates the damage versus excitation time should saturate at any level below 1.0 or rather behave like an asymptote. Once a material undergoes critical or maximum damage while propagating through the captured path, the damage parameter becomes constant with no further release of fracture energy. This phenomenon is well justified by the simulation results obtained in the present study.   
\section{Mesh sensitivity analysis and computational stability of EIPDM and CMDFSI}
Computational analysis using finite element techniques for materials with assigned plasticity damage mechanical model proposed in this study is critically sensitive to accurate mesh sizes. It is indeed very important to assign a near-perfect grid size to capture realistic profiles of principal stresses and fracture paths from the model. Meshes with both coarse and finesse grid sizes can capture stresses and fracture but the major differences lie in the accuracy of these stresses and fracture paths captured. 
\begin{figure*}[!t]
	\captionsetup[subfloat]{justification=centering}
	\subfloat[A CPE4R mesh with global grid to length ratio of 4.2\label{subfig-1:dummy1}]{%
		\includegraphics[width=0.5\textwidth]{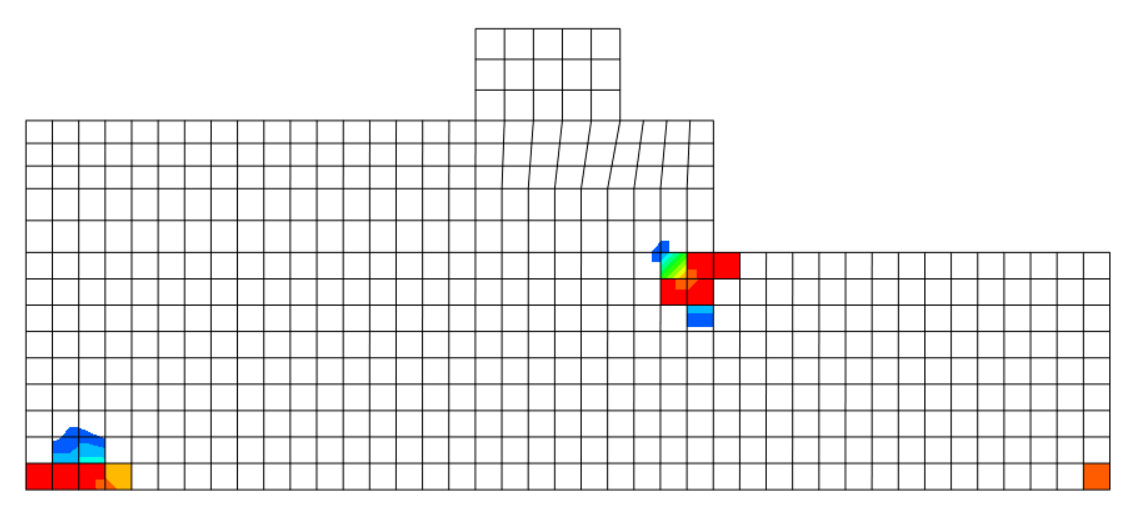}
	}
	\hfill
	\subfloat[A CPE mesh with global grid to length ratio of 4.2\label{subfig-2:dummy2}]{%
		\includegraphics[width=0.5\textwidth]{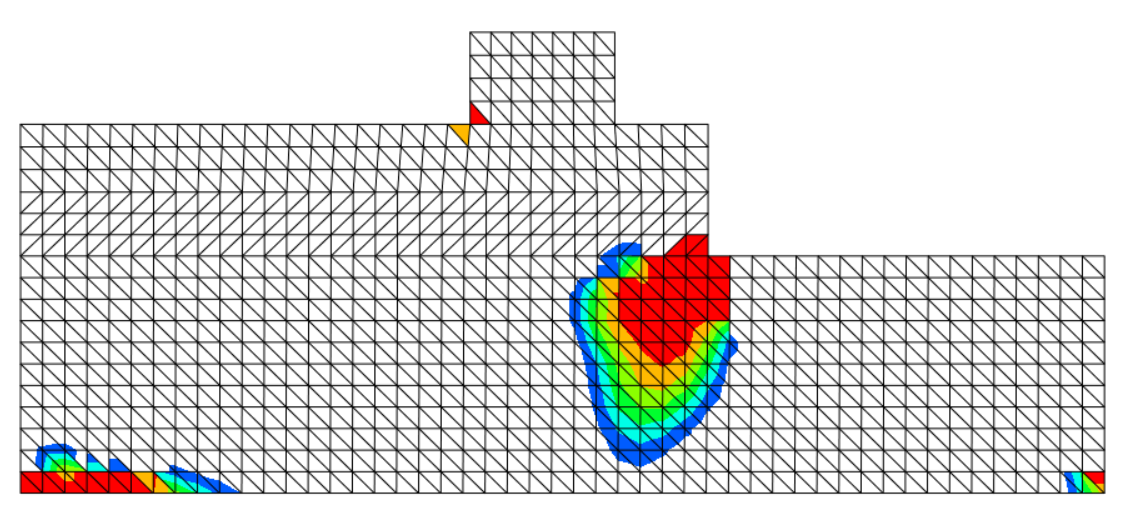}
	}
	\caption{Plane strain computational models showing the tensile damage profile captured by simulation for different meshing types.}
	\label{fig13}
\end{figure*}
Meshes with coarser grid sizes capture discontinuous or rather discrete stress profiles while meshes with finer grid sizes can capture accurate stress profiles having planar continuity. Coarse grid size generates localized regions of stresses that more or less look alike pseudo stress concentrations thus a mesh sensitivity analysis is performed for mesh sizes having a non-dimensional grid to length ratios ranging from 1.0 to 7.0. The computational simulation was performed on both structured CPE4R quad and structured CPE tri meshes.
\begin{figure*}[!t]
	\captionsetup[subfloat]{justification=centering}
	\subfloat[A CPE4R mesh with global grid to length ratio of 4.2\label{subfig-1:dummy}]{%
		\includegraphics[width=0.5\textwidth]{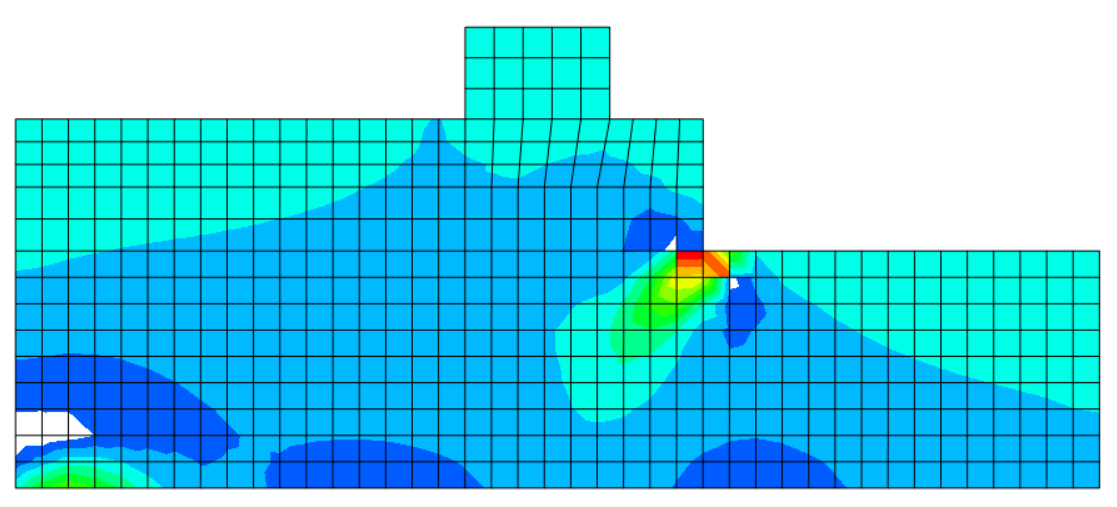}
	}
	\hfill
	\subfloat[A CPE mesh with global grid to length ratio of 2.1\label{subfig-2:dummy}]{%
		\includegraphics[width=0.5\textwidth]{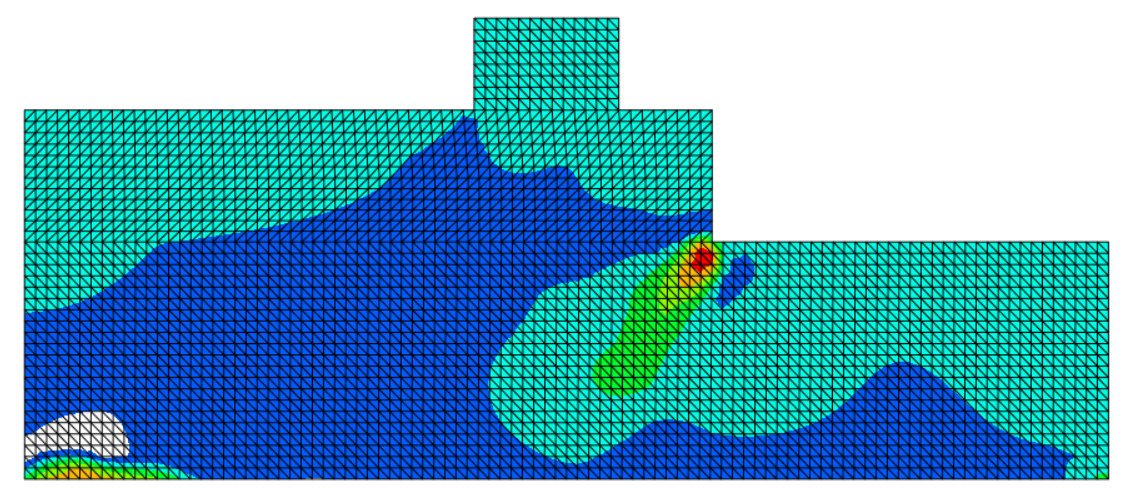}
	}
	\caption{Plane strain computational models showing the maximum principal stresses captured near crack openings for different meshing types.}
	\label{fig14}
\end{figure*}
From \autoref{fig13} it is evident that maximum principal stresses captured by the simulation are distinctive of each other at different levels of mesh finesse and for different types of meshes. For quad dominated CPE4R mesh having a grid to length ratio of 4.2, the critical maximum principle stress spreads discretely through two unique integration points whereas, for a CPE4R quad mesh with a grid length ratio of 2.1, critical maximum principle stress spreads over a continuum region of 7 to 8 unique integration points. For CPE elemental meshes having tri dominated mesh type, the stress profile captured becomes like a stretched ellipse with continuous contour but is more prominent at a higher grid to length ratio. Stretched elliptical profile of stress contour highlights realistic stress flows through the material due to seismic excitation and multiphase fluid-structure interaction. Stress concentration occurs at the location of fracture initiation during computation using only the enhanced version of the plasticity damage mechanical model. When multiphase dynamic fluid-structure interaction is included in the computation along with the enhanced plasticity damage model, the concentrated stresses near the crack opening or fracture initiation region spreads to the adjoining integration points or computational grid, this creates the realistic stretched elliptical stress flow profile as shown in \autoref{subfig-2:dummy}. Finite element computation using structured cosserat point element with three noded triangles gave precise fracture path or rather plane strain yield surface when compared to mesh having four noded quad element. From \autoref{subfig-1:dummy1} and \autoref{subfig-2:dummy2} it is observed that discrete fracture paths were obtained for quad meshing with a grid to length ratio of 4.2 whereas, the fracture surface obtained for tri meshing with the same grid to length ratio was having a near elliptical and continuous profile towards the top. Each and every type of mesh behaves differently with a varying grid to length ratios thus it is very difficult to arrive at a suitable or rather optimized grid to length ratio for obtaining accurate computational results. Based on observations from contours shown in \autoref{fig13} and \autoref{fig14}, three noded CPE type element is adopted for computational stability analysis. Instability in computation arises when the grids inside a mesh are either two closely spaced forming regions of overlap or are too widely stretched. Widely stretched grids inside any mesh domain result in isoparametric distortion which further delays the convergence of equilibrium iterations \citep{Lee1993}.
\begin{figure*}
	\centering
	\includegraphics[width=0.8\textwidth, height=6cm]{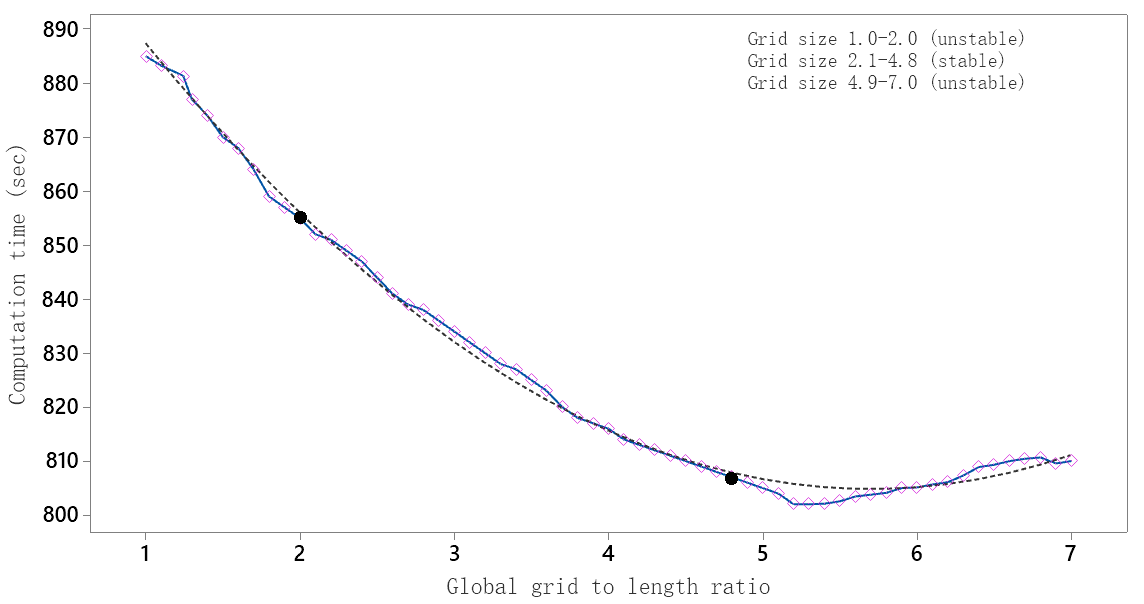}\caption{A mesh sensitivity curve showing the computation time required for each value of global grid to length ratios between 1.0 and 7.0. The curve also highlights the stable and unstable computational regions.}
	\label{fig15}
\end{figure*}
 From \autoref{fig15} it is observed that with an increase in the global grid to length ratio from 1.0 to 5.0, the time required for computational convergence decreases, and post 5.0 the computation time increases. A cubic interpolation curve is fitted to study the stability parameter and it is observed that due to excessive grid overlapping, the computation curve is unstable for a grid to length ratio ranging from 1.0 to 2.0. Post 4.9, the curve becomes unstable yet again due to excessive stretching of the grids inside the mesh. Optimum computational stability is achieved only when the ratio was set between a range of 3.6 to 4.8.
\section{Response of tidal barrage to seismic excitation using EIPDM and CMDFSI }
A three-dimensional model of the tidal barrage present within the enclosure of the Kalpasar dam is simulated using finite element computation in Abaqus to study the fracture propagation due to seismic excitation. The improved damage plasticity mechanical model and idealized Lagrangian-Eulerian multiphase fluid-structure interaction model are adopted for computational simulation. The width and height of the barrage is taken as 205 m and 52 m respectively. The width and height of the discharge channel are taken as 92 m and 14 m respectively. A crested ogee is modeled at the discharge end having a curvature profile of ${x^{1.85}} = 10.01 * y$ towards the downstream monolith and a 1:1 linear slope towards the upstream monolith. The overall length of the barrage is around 1200 m to 1300 m which is beyond the computational scope of this study thus an effective length between two construction joints from the barrage is considered for analysis and the value is taken as 182 m approximately.
\begin{table*}[!t]
	\caption{Material elasticity properties for quasi-brittle concrete of high strength mix-30 as per Indian standards used in plasticity-damage computation.}
	\label{tab3}
	\resizebox{\textwidth}{!}{%
		\begin{tabular}{@{}lcccc@{}}
			\toprule
			\begin{tabular}[c]{@{}l@{}}Strain softening points\\ (Cornelissen Interpolation)\end{tabular} & \begin{tabular}[c]{@{}c@{}}Cracking stress (Pa)\\ $\varepsilon \left( \Im  \right)_{tens}^{^{cracking,h}}$\end{tabular} & \begin{tabular}[c]{@{}c@{}}Crack displacement (m)\\ $\delta \left( \Im  \right)_{tens}^{^{cracking,h}}$\end{tabular} & \begin{tabular}[c]{@{}c@{}}Tensile Damage\\ ${{{\mathord{\buildrel{\lower3pt\hbox{$\scriptscriptstyle\frown$}}\over 
								d} }_{tens}}}$\end{tabular} & \begin{tabular}[c]{@{}c@{}}Improved Tensile Damage\\ ${{\mathord{\buildrel{\lower3pt\hbox{$\scriptscriptstyle\frown$}}\over d} }^{imp}}_{tens}$\end{tabular} \\ \midrule
			0.7*${({f_{ck}})^{0.5}}$ & 3834057.903 & 0 & 0 & 0 \\
			0.467*${({f_{ck}})^{0.5}}$ & 2553482.563 & 2.00E-05 & 0.334 & 0.313 \\
			0.263*${({f_{ck}})^{0.5}}$ & 1437771.713 & 0.00011 & 0.625 & 0.611 \\
			0.117*${({f_{ck}})^{0.5}}$ & 638754.0466 & 0.00021 & 0.8334 & 0.812 \\
			0.0086*${({f_{ck}})^{0.5}}$ & 47144.9958 & 0.0005 & 0.98770363 & 0.9788 \\
			0 & 0 & 0.00056 & 1 & 1 \\ \bottomrule
		\end{tabular}%
	}
\end{table*}
 The width of the Eulerian domain is taken as 50 m without any interaction and an additional 95 m with interaction. The maximum water level reached at the reservoir end within a complete tidal cycle is considered to be the equivalent height of interaction for the Eulerian domain and the value is adopted as 6.99 m. An undeformable rigid plate is assigned to the bottom boundary of the Eulerian domain and the remaining truncated boundaries are assigned with zero degrees of freedom. Only the top surface of the Eulerian domain is allowed to undergo deformation analogous to a free water surface under seismic excitation. The interaction of barrage bottom with it's foundation having rocks and granites with an approximate elastic modulus of 7.17x10\textsuperscript{4} MPa behaves quite rigidly thus neglected in the simulation.
 \begin{figure*}[!t]
 	\centering
 	\includegraphics[width=\textwidth]{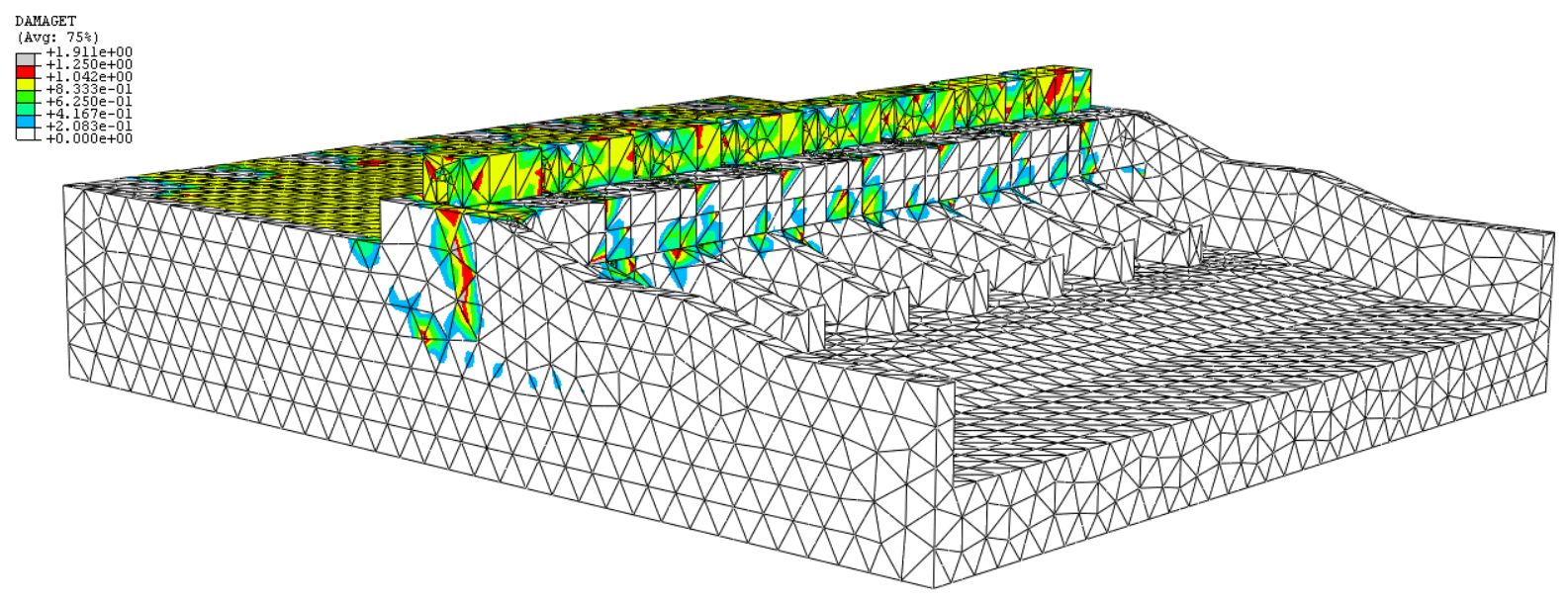}\caption{A three dimensional computational model of concrete tidal barrage highlighting the captured tensile damage and crack growth profile on the left retaining wall and the gate housing.}
 	\label{fig17}
 \end{figure*}
A pseudo volumetric water mass is assigned to the Eulerian domain having a mass density of 1000 kg/m\textsuperscript{3}, bulk modulus of 2060 MPa, and rotational constrain of 1.54x10\textsuperscript{5} MPa. The speed of sound over the free water surface is assigned as 1500 m/s. The enhanched version of the plasticity damage mechanical model proposed in this study with values as given in \autoref{tab3} is adopted. The elastic modulus of M-30 Indian standard mix concrete is considered as 27.386 MPa, Poisson ratio as 0.28, mass density as 2670 kg/m\textsuperscript{3} and yield strength in compression as 30 MPa. The dynamic variation of elastic modulus with time during seismic excitation is neglected. A linear perturbation analysis is performed to extract the modal frequencies of the barrage under seismic excitation. Solving \autoref{eq32} with fundamental frequency as 2.805 and critical damping as 0.003, Rayleigh's stiffness proportional damping constant is obtained with a value of 0.003404. The transverse and vertical seismic loads assigned are extracted from
\begin{figure*}[t!]
	\captionsetup[subfloat]{justification=centering}
	\subfloat[Crack at 2.096 seconds\label{subfig-1:dummy4}]{%
		\includegraphics[width=0.32\textwidth]{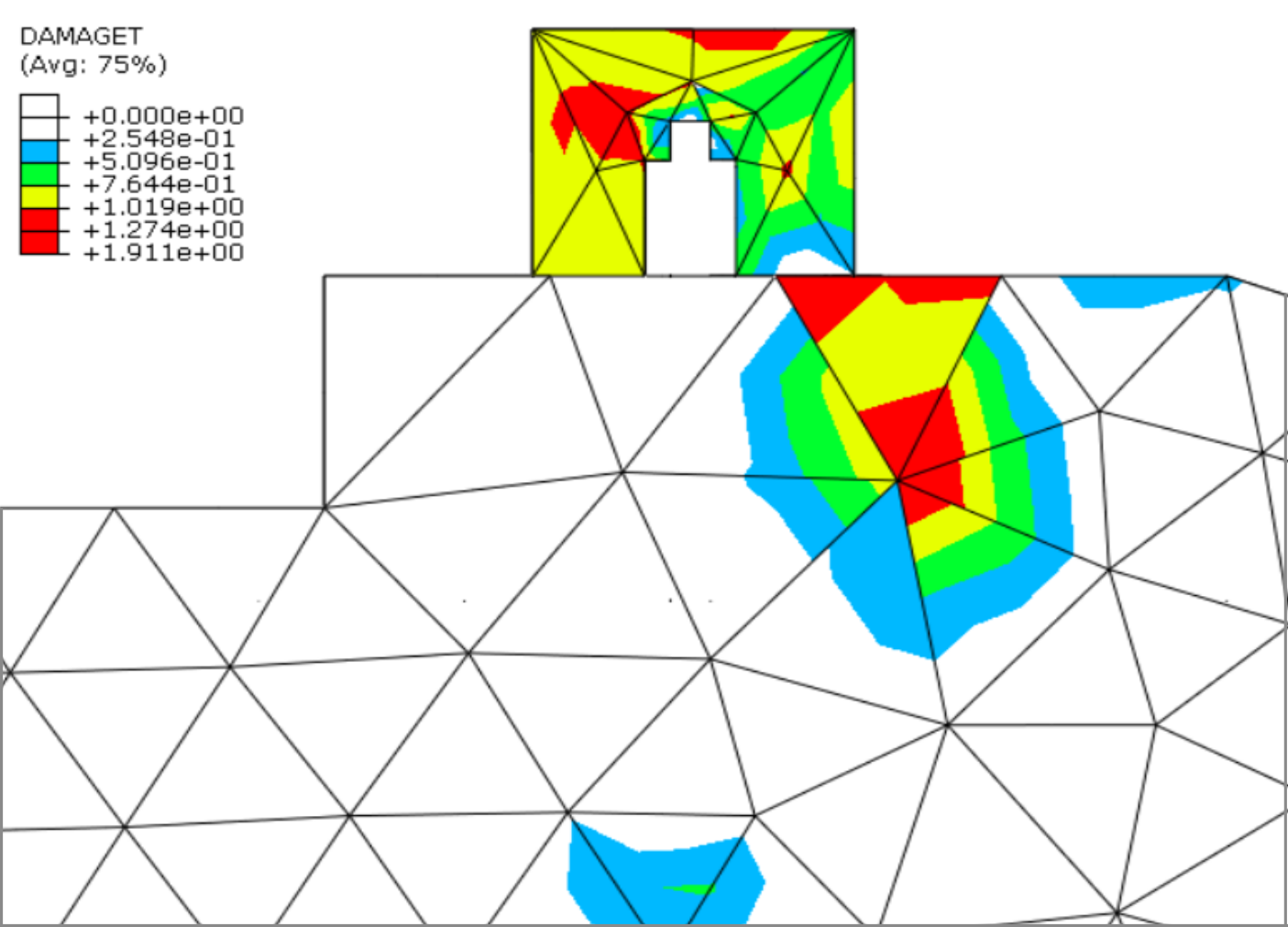}
	}
	\hfill
	\subfloat[Crack at 5.95 seconds\label{subfig-2:dummy4}]{%
		\includegraphics[width=0.32\textwidth]{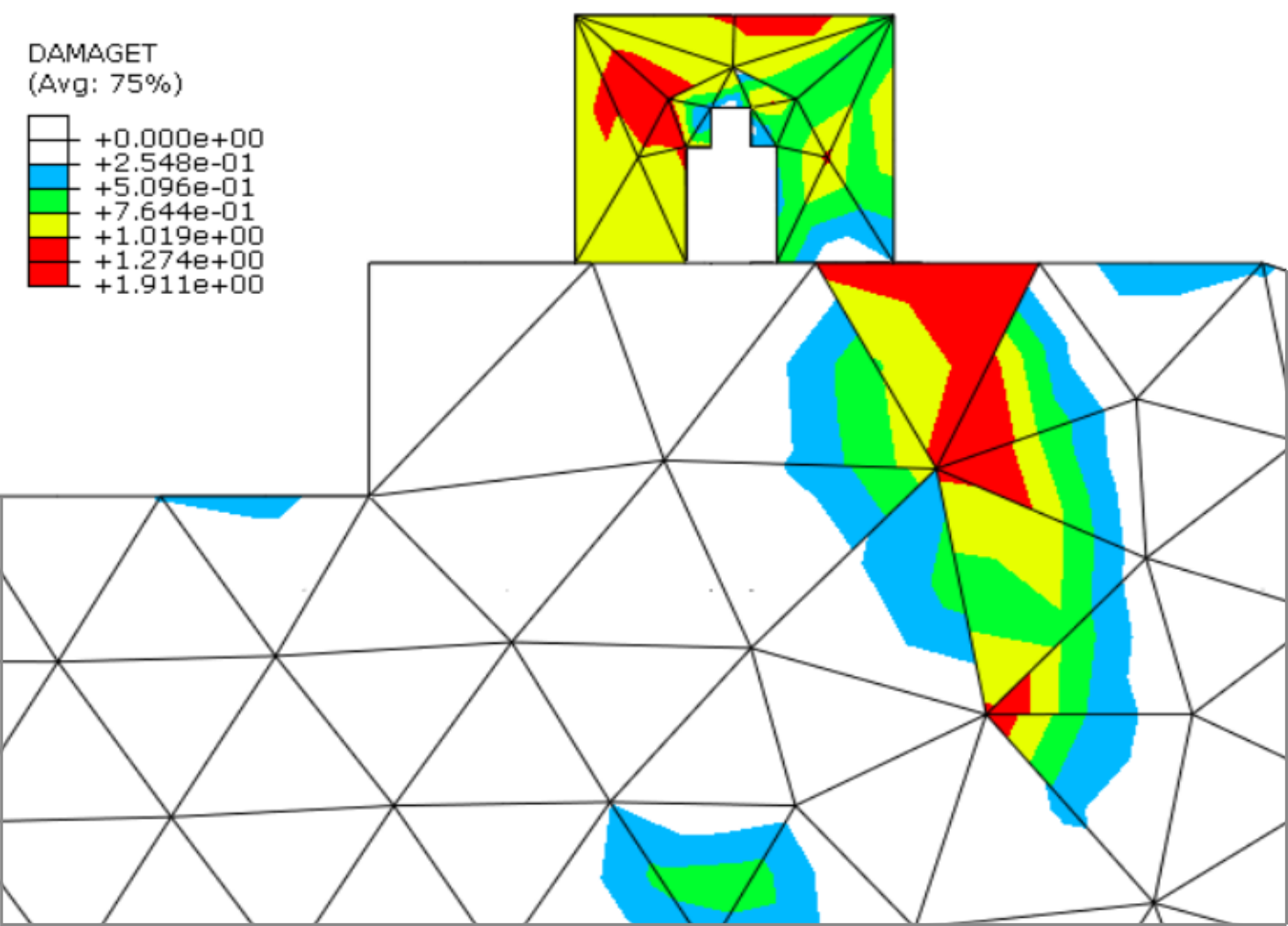}
	}
	\hfill
	\subfloat[Crack at 6.86 seconds\label{subfig-3:dummy4}]{%
		\includegraphics[width=0.32\textwidth]{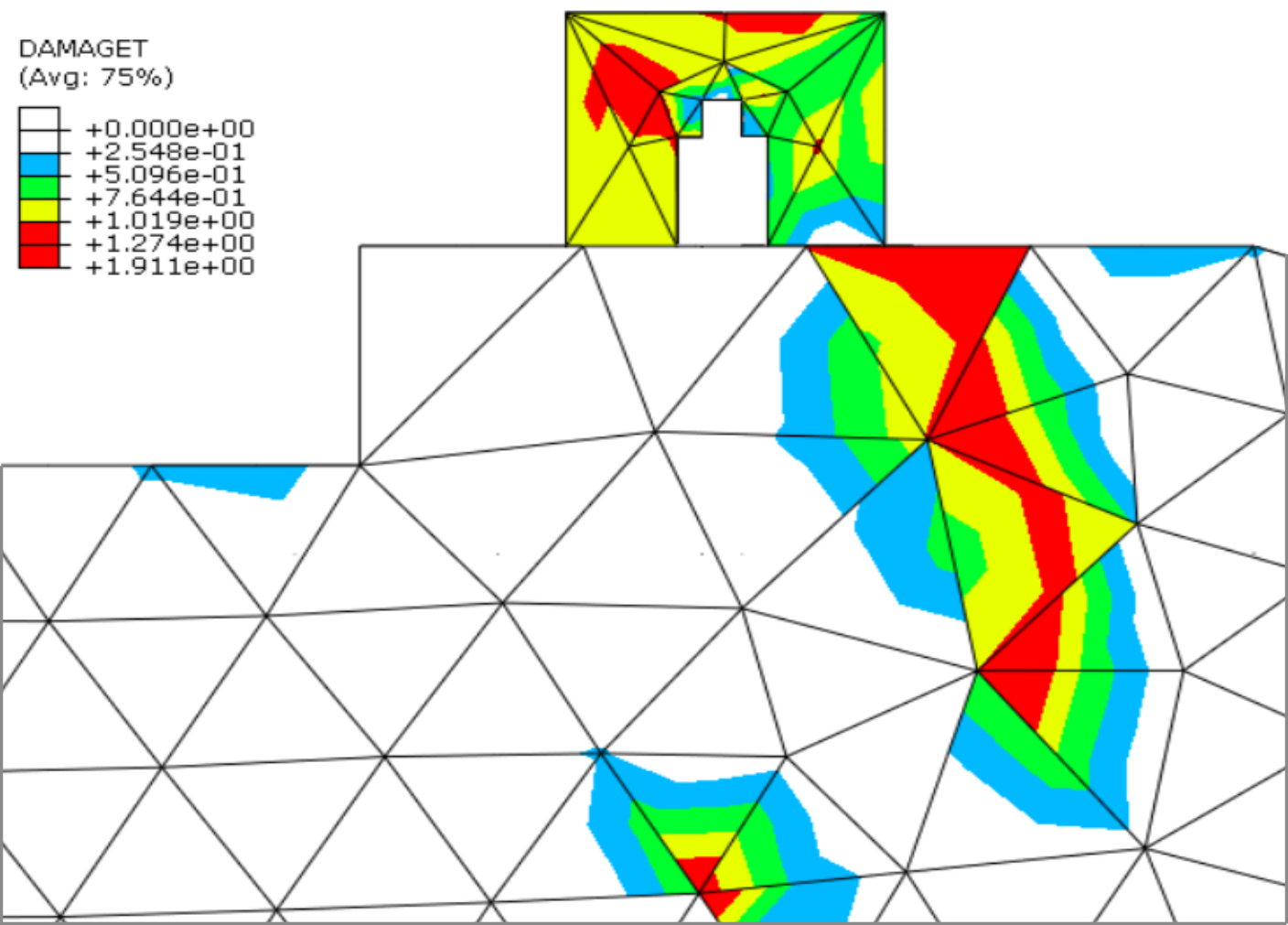}
	}
	\caption{Fracture profile captured on the outer surface of the left retaining wall and crack propagation pattern at different seismic excitation time.}
	\label{fig19}
\end{figure*}
\begin{figure*}[t!]
	\centering
	\includegraphics[width=0.9\textwidth, height=6.5cm]{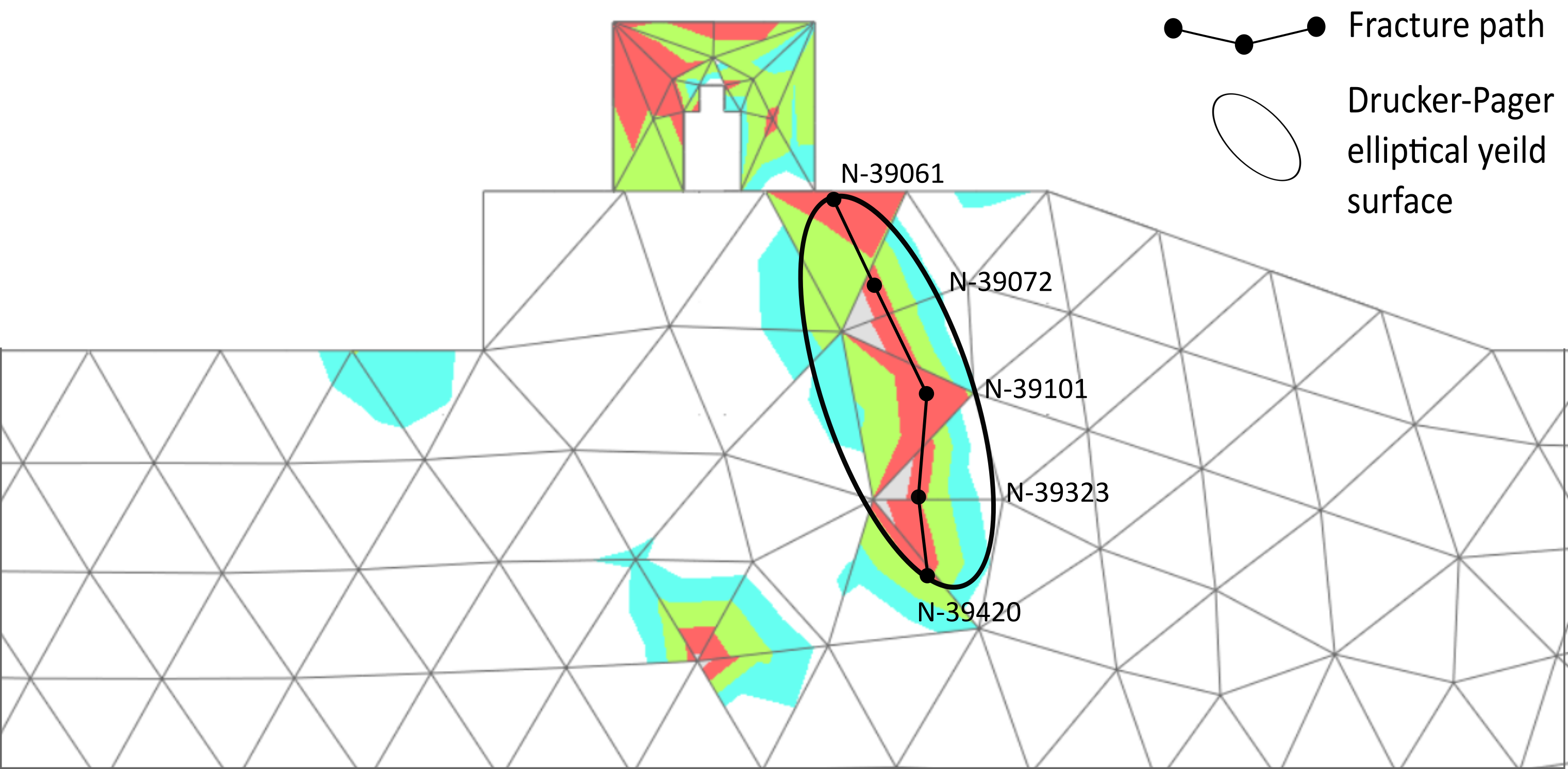}\caption{A planar visualization of the outer surface of left retaining wall showing fracture initiation and propagation through an stretched elliptical Drucker-Pager type yield surface.}
	\label{fig20}
\end{figure*}
 the strong ground motion data recorded for a historic love wave propagating from station ALIR to station GIRNAR directly passing below the Kalpasar dam through the Cambay rift as shown in \autoref{fig:2} having a magnitude of M\textsubscript{w}-6.7 as per Richter scale. The strong ground data is extracted from historic love wave propagating from station ALIR to station GIRNAR then filtered and plotted against an excitation time of 10 seconds. Based on mesh sensitivity and stability computation as shown in \autoref{fig15}, a C3D4H type tetrahedral which is a three-dimensional hybrid form of hex mesh with a global grid to length ratio of 4.2 is adopted for analysis. From \autoref{fig17} it is observed that the damage is mostly localized in the upper region of the barrage with cracks initiating from the gate housing and propagating through the retaining wall up to a height of 15 meters from the top elevation. The majority of the damage occurred at the reservoir end and on the surfaces of both the retaining walls. The fracture profile on the left retaining wall with a stretched elliptical yield surface propagated at an angle of 22.5\textdegree from the vertical while that on the right retaining wall propagated at an angle of 75\textdegree from the vertical. From \autoref{fig17}, it is also observed that the inner surface of the left retaining wall has patches of tensile damage near the bottom lining of the barrage. It is also observed that damage propagates at the intersection of gate housing and training piers.
\begin{figure*}[t!]
	\centering
	\includegraphics[width=0.85\textwidth, height=7cm]{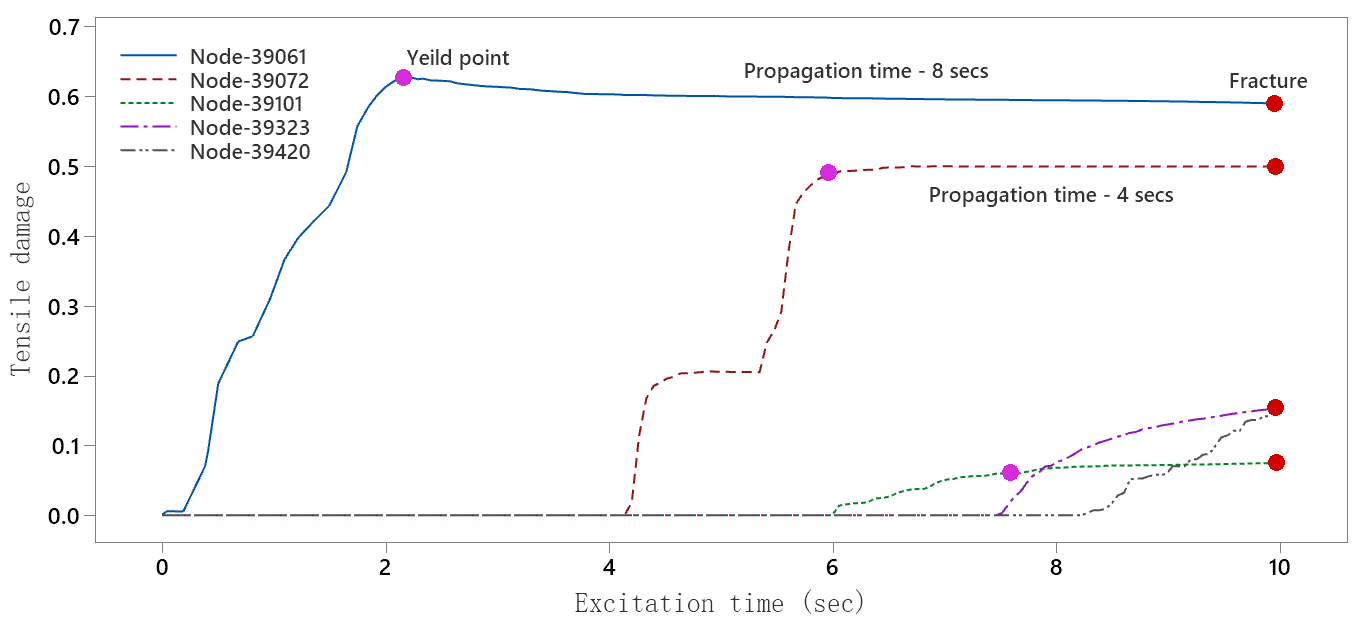}\caption{A tensile damage curve of Kalpasar tidal barrage highlighting the fracture propagation characteristics post material yielding at different unique integration nodes present on the fracture path.}
	\label{fig21}
\end{figure*}
The fracture propagation on the outer surface of the left retaining wall at different excitation times is shown in \autoref{fig19} and it is observed that fracture at any tetrahedral element initiates at the integration point present at one of the apex and propagates towards another apex through triangular planes and is divergent in nature. As the excitation time approaches 10 seconds, the fracture propagation patterns slowly become convergent in nature and stretch out on the yield surface giving a much more realistic profile. The divergent spreading is captured in \autoref{subfig-1:dummy4} whereas the convergent spreading is evident from \autoref{subfig-3:dummy4}. While formulating the enhanced version of the plasticity damage mechanical model, an elliptical Drucker-Pager yield surface defined by a hyperbolic yield function was proposed shown in \autoref{fig:5} and formulated as \autoref{eqs313}. The formulation stands valid as the yield surface captured in the simulation comes out to be a stretched elliptical one with cracks propagating from one end of the major axis towards the other.\autoref{fig20} highlights the elliptical profile captured with crack initiating from node 39061 located at the elevation of +12.20 m from mean sea level and ending at node 39420 located at -3.80 m from the mean sea level. In the initial stage of seismic excitation, it is observed that fracture initiated at node 39061, yields corresponding to a damage value of 0.63, and fracture propagates slowly and steadily for an average time period of 8 seconds. As the excitation time increases, nodes start yielding at damage values lower than 0.63 and the propagation time also decreases from 8 to 4 seconds. At the final phase of the excitation material yields corresponding to damage value as low as 0.1 and failure occurs without any significant fracture propagation as shown in \autoref{fig21}.
\section{Conclusions}
In this study, the fracture and damage response of a concrete tidal barrage due to seismic excitation is investigated based on a coupled multiphase fluid-structure interaction between the barrage, reservoir and the gulf. An improved numerical version of the existing plasticity damage mechanical model is formulated by restoring the additional fracture energy losses. The enhanced isotropic plasticity damage mechanical model is also validated against results obtained from benchmark experiments performed by several researchers on the Koyna Dam. Validation showcased 92\% to 95\% accuracy with realistic fracture growth profiles. Finite element meshing is performed with C3D4H-tetrahedral elements having a global grid to length ratio of 4.2 adopted based on mesh sensitivity and computational stability analysis. Element nodal matrices obtained from the coupled multiphase fluid-structure interaction model are assembled numerically using the Lax-Wendroff temporal marching scheme. The non-linear coupled multiphase dynamic equation is solved using a corrected implicit-$\lambda$ integration scheme and predictor-corrector time marching algorithm based on the fracture mechanics principle. Computation using an uncoupled Lagrangian model captured localized stress and pressure concentration near the fracture path within the yield surface, this additional concentration accelerated the crack growth and the fracture profile obtained was wider and coarse. The idealized multiphase fluid-structure interaction formulated using coupled Lagrangian-Eulerian model is able to distribute or rather eradicate the concentrated pressure and stresses near the crack tip which resulted in the formation of a realistic and stretched fracture path with moderate to fine crack width. Fracture initiates at the bottom of the gate housing and propagates through an elliptical Drucker-Pager yield surface till the mid-region of the retaining wall. The majority of the damage occurs at the outer and inner surface of the retaining wall and also at monoliths connecting piers with gate housing. After material yields, fracture propagates rather slowly during the initiation phase whereas, towards the end of seismic excitation failure and fracture occur simultaneously without sufficient warnings.

\bibliographystyle{elsarticle-harv}
\bibliography{library}
\end{document}